\documentclass[rmp,twocolumn,twoside]{revtex4}

\usepackage{graphicx}

\bibpunct{[}{]}{,}{n}{}{,}

\newcommand{\defn}{\textit}
\newcommand{\half}{\mbox{$\frac12$}}
\renewcommand{\d}{{\rm d}}

\newcommand{\e}{{\rm e}}
\newcommand{\set}[1]{\lbrace#1\rbrace}
\newcommand{\av}[1]{\left\langle#1\right\rangle}
\newcommand{\eref}[1]{(\ref{#1})}
\newcommand{\etal}{{\it{}et~al.}}

\newcommand{\cL}{\mathcal{L}}
\newcommand{\Beta}{\mathrm{B}}

\begin{document}

\title{Power laws, Pareto distributions and Zipf's law}
\author{M. E. J. Newman}
\affiliation{Department of Physics and Center for the Study of Complex
Systems,
University of Michigan, Ann Arbor, MI 48109.  U.S.A.}

\begin{abstract}
When the probability of measuring a particular value of some quantity
varies inversely as a power of that value, the quantity is said to follow a
power law, also known variously as Zipf's law or the Pareto distribution.
Power laws appear widely in physics, biology, earth and planetary sciences,
economics and finance, computer science, demography and the social
sciences.  For instance, the distributions of the sizes of cities,
earthquakes, solar flares, moon craters, wars and people's personal
fortunes all appear to follow power laws.  The origin of power-law
behaviour has been a topic of debate in the scientific community for more
than a century.  Here we review some of the empirical evidence for the
existence of power-law forms and the theories proposed to explain them.
\end{abstract}

\maketitle

\section{Introduction}
\begin{figure*}
\begin{center}
\resizebox{12.5cm}{!}{\includegraphics{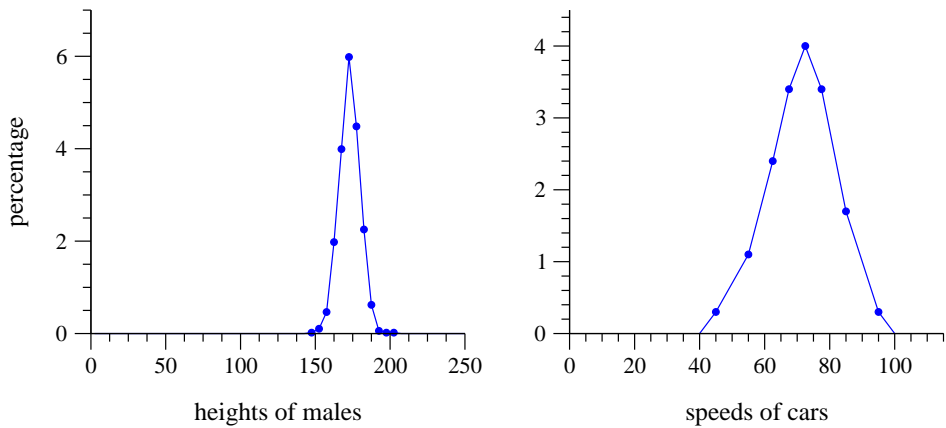}}
\end{center}
\caption{Left: histogram of heights in centimetres of American males.
Data from the National Health Examination Survey, 1959--1962 (US Department
of Health and Human Services).  Right: histogram of speeds in miles per
hour of cars on UK motorways.  Data from Transport Statistics 2003 (UK
Department for Transport).}
\label{scale}
\end{figure*}

Many of the things that scientists measure have a typical size or
``scale''---a typical value around which individual measurements are
centred.  A simple example would be the heights of human beings.  Most
adult human beings are about 180cm tall.  There is some variation around
this figure, notably depending on sex, but we never see people who are 10cm
tall, or 500cm.  To make this observation more quantitative, one can plot a
histogram of people's heights, as I have done in Fig.~\ref{scale}a.  The
figure shows the heights in centimetres of adult men in the United States
measured between 1959 and 1962, and indeed the distribution is relatively
narrow and peaked around 180cm.  Another telling observation is the ratio
of the heights of the tallest and shortest people.  The Guinness Book of
Records claims the world's tallest and shortest adult men (both now dead)
as having had heights 272cm and 57cm respectively, making the ratio 4.8.
This is a relatively low value; as we will see in a moment, some other
quantities have much higher ratios of largest to smallest.

Figure~\ref{scale}b shows another example of a quantity with a typical
scale: the speeds in miles per hour of cars on the motorway.  Again the
histogram of speeds is strongly peaked, in this case around 75mph.

\begin{figure*}
\begin{center}
\resizebox{12cm}{!}{\includegraphics{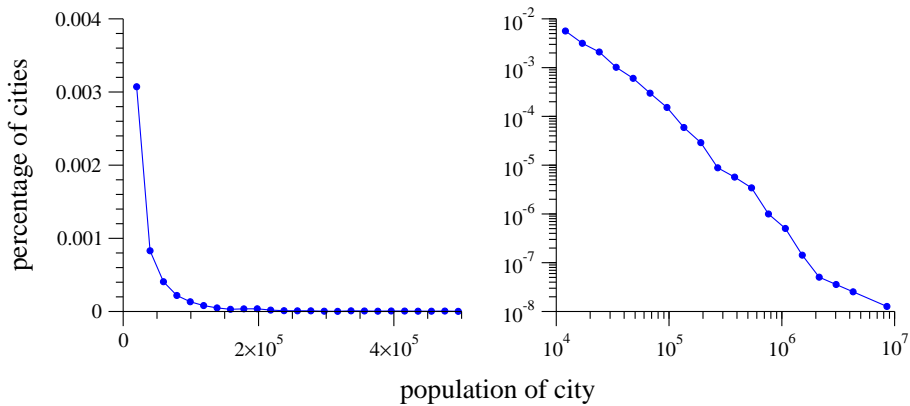}}
\end{center}
\caption{Left: histogram of the populations of all US cities with
population of $10\,000$ or more.  Right: another histogram of the same
data, but plotted on logarithmic scales.  The approximate straight-line
form of the histogram in the right panel implies that the distribution
follows a power law.  Data from the 2000 US Census.}
\label{cities}
\end{figure*}

But not all things we measure are peaked around a typical value.  Some vary
over an enormous dynamic range, sometimes many orders of magnitude.  A
classic example of this type of behaviour is the sizes of towns and cities.
The largest population of any city in the US is 8.00 million for New York
City, as of the most recent (2000) census.  The town with the smallest
population is harder to pin down, since it depends on what you call a town.
The author recalls in 1993 passing through the town of Milliken, Oregon,
population~4, which consisted of one large house occupied by the town's
entire human population, a wooden shack occupied by an extraordinary number
of cats and a very impressive flea market.  According to the Guinness Book,
however, America's smallest town is Duffield, Virginia, with a population
of 52.  Whichever way you look at it, the ratio of largest to smallest
population is at least $150\,000$.  Clearly this is quite different from
what we saw for heights of people.  And an even more startling pattern is
revealed when we look at the histogram of the sizes of cities, which is
shown in Fig.~\ref{cities}.

In the left panel of the figure, I show a simple histogram of the
distribution of US city sizes.  The histogram is highly
\defn{right-skewed}, meaning that while the bulk of the distribution occurs
for fairly small sizes---most US cities have small populations---there is a
small number of cities with population much higher than the typical value,
producing the long tail to the right of the histogram.  This right-skewed
form is qualitatively quite different from the histograms of people's
heights, but is not itself very surprising.  Given that we know there is a
large dynamic range from the smallest to the largest city sizes, we can
immediately deduce that there can only be a small number of very large
cities.  After all, in a country such as America with a total population of
300 million people, you could at most have about 40 cities the size of New
York.  And the 2700 cities in the histogram of Fig.~\ref{cities} cannot
have a mean population of more than $3\times10^8/2700 = 110\,000$.

What is surprising on the other hand, is the right panel of
Fig.~\ref{cities}, which shows the histogram of city sizes again, but this
time replotted with logarithmic horizontal and vertical axes.  Now a
remarkable pattern emerges: the histogram, when plotted in this fashion,
follows quite closely a straight line.  This observation seems first to
have been made by Auerbach~\cite{Auerbach13}, although it is often
attributed to Zipf~\cite{Zipf49}.  What does it mean?  Let $p(x)\>\d x$ be
the fraction of cities with population between~$x$ and $x+\d x$.  If the
histogram is a straight line on log-log scales, then $\ln p(x) = -\alpha\ln
x + c$, where $\alpha$ and $c$ are constants.  (The minus sign is optional,
but convenient since the slope of the line in Fig.~\ref{cities} is clearly
negative.)  Taking the exponential of both sides, this is equivalent to:
\begin{equation}
p(x) = C x^{-\alpha},
\label{powerlaw}
\end{equation}
with $C=\e^c$.

Distributions of the form~\eref{powerlaw} are said to follow a \defn{power
law}.  The constant $\alpha$ is called the \defn{exponent} of the power
law.  (The constant~$C$ is mostly uninteresting; once $\alpha$ is fixed, it
is determined by the requirement that the distribution $p(x)$ sum to~1; see
Section~\ref{normalization}.)

Power-law distributions occur in an extraordinarily diverse range of
phenomena.  In addition to city populations, the sizes of
earthquakes~\cite{GR44}, moon craters~\cite{NI94}, solar
flares~\cite{LH91}, computer files~\cite{CB96} and wars~\cite{RT98}, the
frequency of use of words in any human language~\cite{Estoup16,Zipf49}, the
frequency of occurrence of personal names in most cultures~\cite{ZM01}, the
numbers of papers scientists write~\cite{Lotka26}, the number of citations
received by papers~\cite{Price65}, the number of hits on web
pages~\cite{AH00b}, the sales of books, music recordings and almost every
other branded commodity~\cite{CFC95,KS03}, the numbers of species in
biological taxa~\cite{WY22}, people's annual incomes~\cite{Pareto96} and a
host of other variables all follow power-law distributions.\footnote{Power
laws also occur in many situations other than the statistical distributions
of quantities.  For instance, Newton's famous $1/r^2$ law for gravity has a
power-law form with exponent~$\alpha=2$.  While such laws are certainly
interesting in their own way, they are not the topic of this paper.  Thus,
for instance, there has in recent years been some discussion of the
``allometric'' scaling laws seen in the physiognomy and physiology of
biological organisms~\cite{WBE97}, but since these are not statistical
distributions they will not be discussed here.}

Power-law distributions are the subject of this article.  In the following
sections, I discuss ways of detecting power-law behaviour, give empirical
evidence for power laws in a variety of systems and describe some of the
mechanisms by which power-law behaviour can arise.

\begin{figure*}
\begin{center}
\resizebox{12cm}{!}{\includegraphics{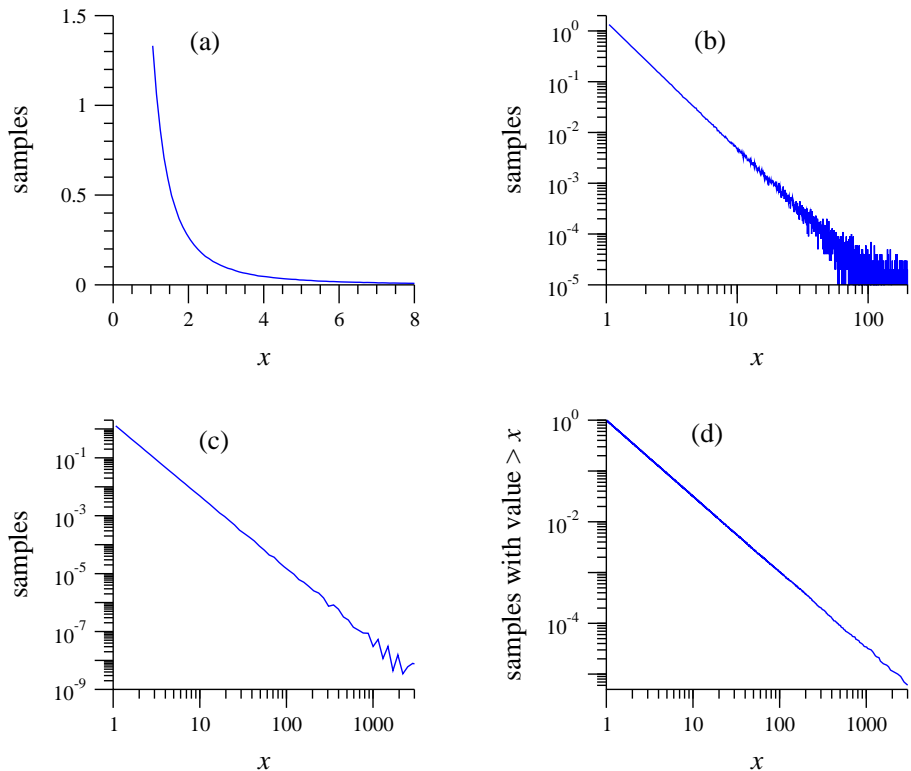}}
\end{center}
\caption{(a)~Histogram of the set of 1 million random
numbers described in the text, which have a power-law distribution with
exponent~$\alpha=2.5$.  (b)~The same histogram on logarithmic scales.
Notice how noisy the results get in the tail towards the right-hand side of
the panel.  This happens because the number of samples in the bins becomes
small and statistical fluctuations are therefore large as a fraction of
sample number.  (c)~A histogram constructed using ``logarithmic binning''.
(d)~A cumulative histogram or rank/frequency plot of the same data.  The
cumulative distribution also follows a power law, but with an exponent of
$\alpha-1=1.5$.}
\label{fake}
\end{figure*}

Readers interested in pursuing the subject further may also wish to consult
the reviews by Sornette~\cite{Sornette03pl} and
Mitzenmacher~\cite{Mitzenmacher04}, as well as the bibliography by
Li.\footnote{\texttt{http://linkage.rockefeller.edu/wli/zipf/}.}

\section{Measuring power laws}
Identifying power-law behaviour in either natural or man-made systems can
be tricky.  The standard strategy makes use of a result we have already
seen: a histogram of a quantity with a power-law distribution appears as a
straight line when plotted on logarithmic scales.  Just making a simple
histogram, however, and plotting it on log scales to see if it looks
straight is, in most cases, a poor way proceed.

Consider Fig.~\ref{fake}.  This example shows a fake data set: I have
generated a million random real numbers drawn from a power-law probability
distribution $p(x)=Cx^{-\alpha}$ with exponent $\alpha=2.5$, just for
illustrative purposes.\footnote{This can be done using the so-called
transformation method.  If we can generate a random real number~$r$
uniformly distributed in the range $0\le r<1$, then
$x=x_\mathrm{min}(1-r)^{-1/(\alpha-1)}$ is a random power-law-distributed
real number in the range $x_\mathrm{min}\le x<\infty$ with
exponent~$\alpha$.  Note that there has to be a lower limit
$x_\mathrm{min}$ on the range; the power-law distribution diverges as
$x\to0$---see Section~\ref{secexamples}.}  Panel~(a) of the figure shows a
normal histogram of the numbers, produced by binning them into bins of
equal size~$0.1$.  That is, the first bin goes from $1$ to~$1.1$, the
second from $1.1$ to~$1.2$, and so forth.  On the linear scales used this
produces a nice smooth curve.

To reveal the power-law form of the distribution it is better, as we have
seen, to plot the histogram on logarithmic scales, and when we do this for
the current data we see the characteristic straight-line form of the
power-law distribution, Fig.~\ref{fake}b.  However, the plot is in some
respects not a very good one.  In particular the right-hand end of the
distribution is noisy because of sampling errors.  The power-law
distribution dwindles in this region, meaning that each bin only has a few
samples in it, if any.  So the fractional fluctuations in the bin counts
are large and this appears as a noisy curve on the plot.  One way to deal
with this would be simply to throw out the data in the tail of the curve.
But there is often useful information in those data and furthermore, as we
will see in Section~\ref{secexamples}, many distributions follow a power
law \emph{only} in the tail, so we are in danger of throwing out the baby
with the bathwater.

An alternative solution is to vary the width of the bins in the histogram.
If we are going to do this, we must also normalize the sample counts by the
width of the bins they fall in.  That is, the number of samples in a bin of
width~$\Delta x$ should be divided by $\Delta x$ to get a count \emph{per
unit interval} of~$x$.  Then the normalized sample count becomes
independent of bin width on average and we are free to vary the bin widths
as we like.  The most common choice is to create bins such that each is a
fixed multiple wider than the one before it.  This is known as
\defn{logarithmic binning}.  For the present example, for instance, we might
choose a multiplier of~2 and create bins that span the intervals 1 to
$1.1$, $1.1$ to $1.3$, $1.3$ to $1.7$ and so forth (i.e.,~the sizes of the
bins are $0.1$, $0.2$, $0.4$ and so forth).  This means the bins in the
tail of the distribution get more samples than they would if bin sizes were
fixed, and this reduces the statistical errors in the tail.  It also has
the nice side-effect that the bins appear to be of constant width when we
plot the histogram on log scales.

I used logarithmic binning in the construction of Fig.~\ref{cities}b, which
is why the points representing the individual bins appear equally spaced.
In Fig.~\ref{fake}c I have done the same for our computer-generated
power-law data.  As we can see, the straight-line power-law form of the
histogram is now much clearer and can be seen to extend for at least a
decade further than was apparent in Fig.~\ref{fake}b.

Even with logarithmic binning there is still some noise in the tail,
although it is sharply decreased.  Suppose the bottom of the lowest bin is
at $x_\mathrm{min}$ and the ratio of the widths of successive bins is~$a$.
Then the $k$th bin extends from $x_{k-1}=x_\mathrm{min} a^{k-1}$ to
$x_k=x_\mathrm{min} a^k$ and the expected number of samples falling in this
interval is
\begin{eqnarray}
\int_{x_{k-1}}^{x_k} \! p(x)\>\d x
    &=& C \int_{x_{k-1}}^{x_k} \! x^{-\alpha} \>\d x \nonumber\\
    &=& C\,{a^{\alpha-1}-1\over\alpha-1} (x_\mathrm{min} a^k)^{-\alpha+1}.
\end{eqnarray}
Thus, so long as $\alpha>1$, the number of samples per bin goes down as $k$
increases and the bins in the tail will have more statistical noise than
those that precede them.  As we will see in the next section, most
power-law distributions occurring in nature have $2\le\alpha\le3$, so noisy
tails are the norm.

Another, and in many ways a superior, method of plotting the data is to
calculate a \defn{cumulative distribution function}.  Instead of plotting a
simple histogram of the data, we make a plot of the probability~$P(x)$ that
$x$ has a value greater than or equal to~$x$:
\begin{equation}
P(x) = \int_x^\infty p(x') \>\d x'.
\label{cumulative}
\end{equation}
The plot we get is no longer a simple representation of the distribution of
the data, but it is useful nonetheless.  If the distribution follows a
power law $p(x)=Cx^{-\alpha}$, then
\begin{equation}
P(x) = C \int_x^\infty {x'}^{-\alpha} \>\d x'
     = {C\over\alpha-1} x^{-(\alpha-1)}.
\end{equation}
Thus the cumulative distribution function~$P(x)$ also follows a power law,
but with a different exponent $\alpha-1$, which is 1 less than the original
exponent.  Thus, if we plot $P(x)$ on logarithmic scales we should again
get a straight line, but with a shallower slope.

But notice that there is no need to bin the data at all to
calculate~$P(x)$.  By its definition, $P(x)$~is well-defined for every
value of~$x$ and so can be plotted as a perfectly normal function without
binning.  This avoids all questions about what sizes the bins should be.
It also makes much better use of the data: binning of data lumps all
samples within a given range together into the same bin and so throws out
any information that was contained in the individual values of the samples
within that range.  Cumulative distributions don't throw away any
information; it's all there in the plot.

Figure~\ref{fake}d shows our computer-generated power-law data as a
cumulative distribution, and indeed we again see the tell-tale
straight-line form of the power law, but with a shallower slope than
before.  Cumulative distributions like this are sometimes also called
\defn{rank/frequency plots} for reasons explained in
Appendix~\ref{rfappendix}.  Cumulative distributions with a power-law form
are sometimes said to follow \defn{Zipf's law} or a \defn{Pareto
distribution}, after two early researchers who championed their study.
Since power-law cumulative distributions imply a power-law form for~$p(x)$,
``Zipf's law'' and ``Pareto distribution'' are effectively synonymous with
``power-law distribution''.  (Zipf's law and the Pareto distribution differ
from one another in the way the cumulative distribution is plotted---Zipf
made his plots with $x$ on the horizontal axis and $P(x)$ on the vertical
one; Pareto did it the other way around.  This causes much confusion in the
literature, but the data depicted in the plots are of course
identical.\footnote{See
\texttt{http://www.hpl.hp.com/research/idl/papers/ranking/} for a useful
discussion of these and related points.})

We know the value of the exponent $\alpha$ for our artificial data set
since it was generated deliberately to have a particular value, but in
practical situations we would often like to estimate~$\alpha$ from observed
data.  One way to do this would be to fit the slope of the line in plots
like Figs.~\ref{fake}b, c or~d, and this is the most commonly used method.
Unfortunately, it is known to introduce systematic biases into the value of
the exponent~\cite{GMY04}, so it should not be relied upon.  For example,
a least-squares fit of a straight line to Fig.~\ref{fake}b gives
$\alpha=2.26\pm0.02$, which is clearly incompatible with the known value of
$\alpha=2.5$ from which the data were generated.

An alternative, simple and reliable method for extracting the exponent is
to employ the formula
\begin{equation}
\alpha = 1 + n \Biggl[ \sum_{i=1}^n \ln {x_i\over x_\mathrm{min}} \Biggr]^{-1}.
\label{mle}
\end{equation}
Here the quantities $x_i$, $i=1\ldots n$ are the measured values of~$x$ and
$x_\mathrm{min}$ is again the minimum value of~$x$.  (As discussed in the
following section, in practical situations $x_\mathrm{min}$~usually
corresponds not to the smallest value of $x$ measured but to the smallest
for which the power-law behaviour holds.)  An estimate of the expected
statistical error~$\sigma$ on~\eref{mle} is given by
\begin{equation}
\sigma =
  \sqrt{n} \Biggl[ \sum_{i=1}^n \ln {x_i\over x_\mathrm{min}} \Biggr]^{-1}
  = {\alpha-1\over\sqrt{n}}.
\label{mleerror}
\end{equation}
The derivation of both these formulas is given in Appendix~\ref{mlmethod}.

Applying Eqs.~\eref{mle} and~\eref{mleerror} to our present data gives an
estimate of $\alpha=2.500\pm0.002$ for the exponent, which agrees well with
the known value of $2.5$.

\begin{figure*}[p]
\begin{center}
\resizebox{13.5cm}{!}{\includegraphics{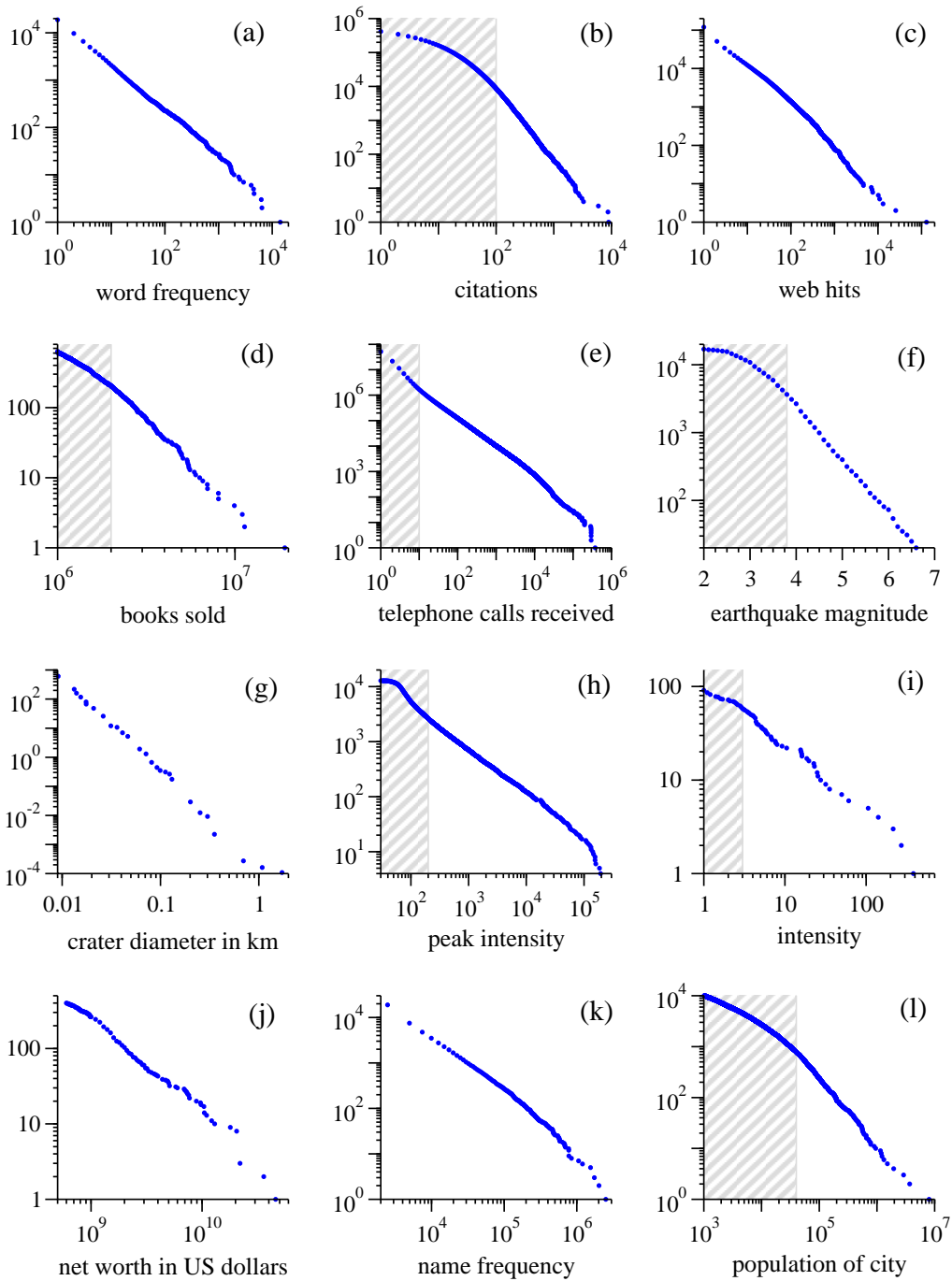}}
\end{center}
\caption{Cumulative distributions or ``rank/frequency plots'' of twelve
quantities reputed to follow power laws.  The distributions were computed
as described in Appendix~\ref{rfappendix}.  Data in the shaded regions were
excluded from the calculations of the exponents in Table~\ref{exponents}.
Source references for the data are given in the text.  (a)~Numbers of
occurrences of words in the novel \textit{Moby Dick} by Hermann Melville.
(b)~Numbers of citations to scientific papers published in 1981, from time
of publication until June 1997.  (c)~Numbers of hits on web sites by
$60\,000$ users of the America Online Internet service for the day of
1~December 1997.  (d)~Numbers of copies of bestselling books sold in the US
between 1895 and 1965.  (e)~Number of calls received by AT\&T telephone
customers in the US for a single day.  (f)~Magnitude of earthquakes in
California between January 1910 and May 1992.  Magnitude is proportional to
the logarithm of the maximum amplitude of the earthquake, and hence the
distribution obeys a power law even though the horizontal axis is linear.
(g)~Diameter of craters on the moon.  Vertical axis is measured per square
kilometre.  (h)~Peak gamma-ray intensity of solar flares in counts per
second, measured from Earth orbit between February 1980 and November 1989.
(i)~Intensity of wars from 1816 to 1980, measured as battle deaths per
$10\,000$ of the population of the participating countries.  (j)~Aggregate
net worth in dollars of the richest individuals in the US in October 2003.
(k)~Frequency of occurrence of family names in the US in the year 1990.
(l)~Populations of US cities in the year 2000.}
\label{examples}
\end{figure*}

\subsection{Examples of power laws}
\label{secexamples}
In Fig.~\ref{examples} we show cumulative distributions of twelve different
quantities measured in physical, biological, technological and social
systems of various kinds.  All have been proposed to follow power laws over
some part of their range.  The ubiquity of power-law behaviour in the
natural world has led many scientists to wonder whether there is a single,
simple, underlying mechanism linking all these different systems together.
Several candidates for such mechanisms have been proposed, going by names
like ``self-organized criticality'' and ``highly optimized tolerance''.
However, the conventional wisdom is that there are actually many different
mechanisms for producing power laws and that different ones are applicable
to different cases.  We discuss these points further in
Section~\ref{generation}.

The distributions shown in Fig.~\ref{examples} are as follows.
\begin{enumerate}
\renewcommand{\theenumi}{\alph{enumi}}
\renewcommand{\labelenumi}{(\theenumi)}
\item \textbf{Word frequency:} Estoup~\cite{Estoup16} observed that the
frequency with which words are used appears to follow a power law, and this
observation was famously examined in depth and confirmed by
Zipf~\cite{Zipf49}.  Panel~(a) of Fig.~\ref{examples} shows the cumulative
distribution of the number of times that words occur in a typical piece of
English text, in this case the text of the novel \textit{Moby Dick} by
Herman Melville.\footnote{The most common words in this case are, in order,
``the'', ``of'', ``and'', ``a'' and ``to'', and the same is true for most
written English texts.  Interestingly, however, it is not true for spoken
English.  The most common words in spoken English are, in order, ``I'',
``and'', ``the'', ``to'' and ``that''~\cite{Dahl79}.}  Similar
distributions are seen for words in other languages.
\item \textbf{Citations of scientific papers:} As first observed by
Price~\cite{Price65}, the numbers of citations received by scientific
papers appear to have a power-law distribution.  The data in panel~(b) are
taken from the Science Citation Index, as collated by
Redner~\cite{Redner98}, and are for papers published in 1981.  The plot
shows the cumulative distribution of the number of citations received by a
paper between publication and June 1997.
\item \textbf{Web hits:} The cumulative distribution of the number of
``hits'' received by web sites (i.e.,~servers, not pages) during a single
day from a subset of the users of the AOL Internet service.  The site with
the most hits, by a long way, was \texttt{yahoo.com}.  After Adamic and
Huberman~\cite{AH00b}.
\item \textbf{Copies of books sold:} The cumulative distribution of the
total number of copies sold in America of the 633 bestselling books that
sold 2 million or more copies between 1895 and 1965.  The data were
compiled painstakingly over a period of several decades by Alice Hackett,
an editor at \textit{Publisher's Weekly}~\cite{Hackett67}.  The best
selling book during the period covered was Benjamin Spock's \textit{The
Common Sense Book of Baby and Child Care}.  (The Bible, which certainly
sold more copies, is not really a single book, but exists in many different
translations, versions and publications, and was excluded by Hackett from
her statistics.)  Substantially better data on book sales than Hackett's
are now available from operations such as Nielsen BookScan, but
unfortunately at a price this author cannot afford.  I should be very
interested to see a plot of sales figures from such a modern source.
\item \textbf{Telephone calls:} The cumulative distribution of the number
of calls received on a single day by 51 million users of AT\&T long
distance telephone service in the United States.  After
Aiello~\etal~\cite{ACL00}.  The largest number of calls received by a
customer in that day was $375\,746$, or about 260 calls a minute (obviously
to a telephone number that has many people manning the phones).  Similar
distributions are seen for the number of calls placed by users and also for
the numbers of email messages that people send and
receive~\cite{EMB02,HA04}.
\item \textbf{Magnitude of earthquakes:} The cumulative distribution of the
Richter (local) magnitude of earthquakes occurring in California between
January 1910 and May 1992, as recorded in the Berkeley Earthquake Catalog.
The Richter magnitude is defined as the logarithm, base 10, of the maximum
amplitude of motion detected in the earthquake, and hence the horizontal
scale in the plot, which is drawn as linear, is in effect a logarithmic
scale of amplitude.  The power law relationship in the earthquake
distribution is thus a relationship between amplitude and frequency of
occurrence.  The data are from the National Geophysical Data Center,
\texttt{www.ngdc.noaa.gov}.
\item \textbf{Diameter of moon craters:} The cumulative distribution of the
diameter of moon craters.  Rather than measuring the (integer) number of
craters of a given size on the whole surface of the moon, the vertical axis
is normalized to measure number of craters per square kilometre, which is
why the axis goes below~1, unlike the rest of the plots, since it is
entirely possible for there to be less than one crater of a given size per
square kilometre.  After Neukum and Ivanov~\cite{NI94}.
\item \textbf{Intensity of solar flares:} The cumulative distribution of
the peak gamma-ray intensity of solar flares.  The observations were made
between 1980 and 1989 by the instrument known as the Hard X-Ray Burst
Spectrometer aboard the Solar Maximum Mission satellite launched in 1980.
The spectrometer used a CsI scintillation detector to measure gamma-rays
from solar flares and the horizontal axis in the figure is calibrated in
terms of scintillation counts per second from this detector.  The data are
from the NASA Goddard Space Flight Center,
\texttt{umbra.nascom.nasa.gov/smm/hxrbs.html}.  See also Lu and
Hamilton~\cite{LH91}.
\item \textbf{Intensity of wars:} The cumulative distribution of the
intensity of 119 wars from 1816 to 1980.  Intensity is defined by taking
the number of battle deaths among all participant countries in a war,
dividing by the total combined populations of the countries and multiplying
by $10\,000$.  For instance, the intensities of the First and Second World
Wars were $141.5$ and $106.3$ battle deaths per $10\,000$ respectively.
The worst war of the period covered was the small but horrifically
destructive Paraguay-Bolivia war of 1932--1935 with an intensity of
$382.4$.  The data are from Small and Singer~\cite{SS82}.  See also Roberts
and Turcotte~\cite{RT98}.
\item \textbf{Wealth of the richest people:} The cumulative distribution of
the total wealth of the richest people in the United States.  Wealth is
defined as aggregate net worth, i.e.,~total value in dollars at current
market prices of all an individual's holdings, minus their debts.  For
instance, when the data were compiled in 2003, America's richest person,
William H. Gates III, had an aggregate net worth of \$46 billion, much of
it in the form of stocks of the company he founded, Microsoft Corporation.
Note that net worth doesn't actually correspond to the amount of money
individuals could spend if they wanted to: if Bill Gates were to sell all
his Microsoft stock, for instance, or otherwise divest himself of any
significant portion of it, it would certainly depress the stock price.  The
data are from
\textit{Forbes} magazine, 6 October 2003.
\item \textbf{Frequencies of family names:} Cumulative distribution of the
frequency of occurrence in the US of the $89\,000$ most common family
names, as recorded by the US Census Bureau in 1990.  Similar distributions
are observed for names in some other cultures as well (for example in
Japan~\cite{MLNM00}) but not in all cases.  Korean family names for
instance appear to have an exponential distribution~\cite{KP04}.
\item \textbf{Populations of cities:} Cumulative distribution of the size of
the human populations of US cities as recorded by the US Census Bureau in
2000.
\end{enumerate}

Few real-world distributions follow a power law over their entire range,
and in particular not for smaller values of the variable being measured.
As pointed out in the previous section, for any positive value of the
exponent~$\alpha$ the function $p(x)=Cx^{-\alpha}$ diverges as $x\to0$.  In
reality therefore, the distribution must deviate from the power-law form
below some minimum value~$x_\mathrm{min}$.  In our computer-generated
example of the last section we simply cut off the distribution altogether
below~$x_\mathrm{min}$ so that $p(x)=0$ in this region, but most real-world
examples are not that abrupt.  Figure~\ref{examples} shows distributions
with a variety of behaviours for small values of the variable measured; the
straight-line power-law form asserts itself only for the higher values.
Thus one often hears it said that the distribution of such-and-such a
quantity ``has a power-law tail''.

Extracting a value for the exponent $\alpha$ from distributions like these
can be a little tricky, since it requires us to make a judgement, sometimes
imprecise, about the value~$x_\mathrm{min}$ above which the distribution
follows the power law.  Once this judgement is made, however, $\alpha$~can
be calculated simply from Eq.~\eref{mle}.\footnote{Sometimes the tail is
also cut off because there is, for one reason or another, a limit on the
largest value that may occur.  An example is the finite-size effects found
in critical phenomena---see Section~\ref{critphen}.  In this case,
Eq.~\eref{mle} must be modified~\cite{GMY04}.}  (Care must be taken to use
the correct value of~$n$ in the formula; $n$~is the number of samples that
actually go into the calculation, excluding those with values
below~$x_\mathrm{min}$, not the overall total number of samples.)

Table~\ref{exponents} lists the estimated exponents for each of the
distributions of Fig.~\ref{examples}, along with standard errors and also
the values of $x_\mathrm{min}$ used in the calculations.  Note that the
quoted errors correspond only to the statistical sampling error in the
estimation of~$\alpha$; they include no estimate of any errors introduced
by the fact that a single power-law function may not be a good model for
the data in some cases or for variation of the estimates with the value
chosen for~$x_\mathrm{min}$.

\begin{table}
\begin{tabular}{ll|cc}
 &         & minimum          & exponent \\
 & quantity & $x_\mathrm{min}$ & $\alpha$ \\
\hline

(a) & frequency of use of words       & 1             & $2.20(1)$ \\

(b) & number of citations to papers   & 100           & $3.04(2)$ \\

(c) & number of hits on web sites     & 1             & $2.40(1)$  \\

(d) & copies of books sold in the US  & $2\,000\,000$ & $3.51(16)$ \\

(e) & telephone calls received        & 10            & $2.22(1)$  \\

(f) & magnitude of earthquakes        & $3.8$         & $3.04(4)$ \\

(g) & diameter of moon craters        & $0.01$        & $3.14(5)$ \\

(h) & intensity of solar flares       & 200           & $1.83(2)$ \\

(i) & intensity of wars               & 3             & $1.80(9)$ \\

(j) & net worth of Americans          & \$600m        & $2.09(4)$ \\

(k) & frequency of family names       & $10\,000$     & $1.94(1)$ \\

(l) & population of US cities         & $40\,000$     & $2.30(5)$

\end{tabular}
\caption{Parameters for the distributions shown in Fig.~\ref{examples}.
The labels on the left refer to the panels in the figure.  Exponent values
were calculated using the maximum likelihood method of Eq.~\eref{mle} and
Appendix~\ref{mlmethod}, except for the moon craters (g), for which only
cumulative data were available.  For this case the exponent quoted is from
a simple least-squares fit and should be treated with caution.  Numbers in
parentheses give the standard error on the trailing figures.}
\label{exponents}
\end{table}

In the author's opinion, the identification of some of the distributions in
Fig.~\ref{examples} as following power laws should be considered
unconfirmed.  While the power law seems to be an excellent model for most
of the data sets depicted, a tenable case could be made that the
distributions of web hits and family names might have two different
power-law regimes with slightly different exponents.\footnote{Significantly
more tenuous claims to power-law behaviour for other quantities have
appeared elsewhere in the literature, for instance in the discussion of the
distribution of the sizes of electrical blackouts~\cite{CTP01,CNBP01}.
These however I consider insufficiently substantiated for inclusion in the
present work.\label{note1}} And the data for the numbers of copies of books
sold cover rather a small range---little more than one decade horizontally.
Nonetheless, one can, without stretching the interpretation of the data
unreasonably, claim that power-law distributions have been observed in
language, demography, commerce, information and computer sciences, geology,
physics and astronomy, and this on its own is an extraordinary statement.

\subsection{Distributions that do not follow a power law}
\label{secnonpl}
\begin{figure}
\begin{center}
\resizebox{8cm}{!}{\includegraphics{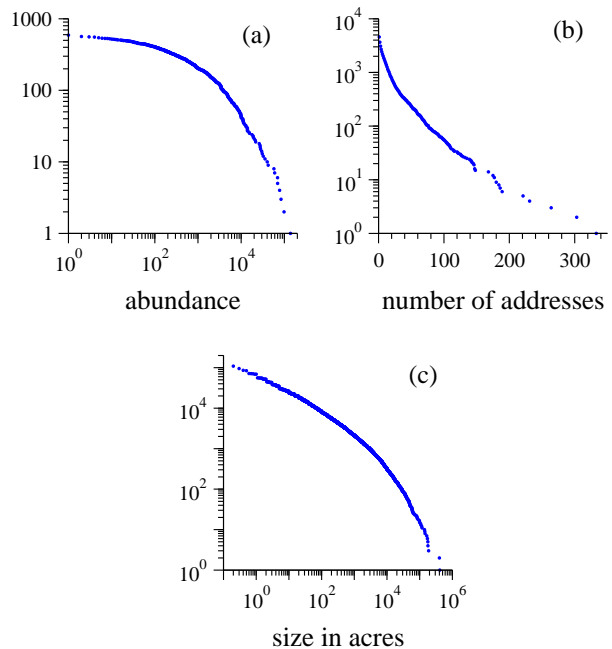}}
\end{center}
\caption{Cumulative distributions of some quantities whose distributions
span several orders of magnitude but that nonetheless do not follow power
laws.  (a)~The number of sightings of 591 species of birds in the North
American Breeding Bird Survey 2003.  (b)~The number of addresses in the
email address books of $16\,881$ users of a large university computer
system \citep{NFB02}.  (c)~The size in acres of all wildfires occurring on
US federal land between 1986 and 1996 (National Fire Occurrence Database,
USDA Forest Service and Department of the Interior).  Note that the
horizontal axis is logarithmic in frames (a) and (c) but linear in
frame~(b).}
\label{nonpl}
\end{figure}

Power-law distributions are, as we have seen, impressively ubiquitous, but
they are not the only form of broad distribution.  Lest I give the
impression that everything interesting follows a power law, let me
emphasize that there are quite a number of quantities with highly
right-skewed distributions that nonetheless do not obey power laws.  A few
of them, shown in Fig.~\ref{nonpl}, are the following:
\begin{enumerate}
\renewcommand{\theenumi}{\alph{enumi}}
\renewcommand{\labelenumi}{(\theenumi)}
\item The abundance of North American bird species, which spans over five
orders of magnitude but is probably distributed according to a log-normal.
A log-normally distributed quantity is one whose logarithm is normally
distributed; see Section~\ref{sundry} and Ref.~\cite{LSA01} for further
discussions.
\item The number of entries in people's email address books, which spans
about three orders of magnitude but seems to follow a stretched
exponential.  A stretched exponential is curve of the form $\e^{-ax^b}$ for
some constants $a, b$.
\item The distribution of the sizes of forest fires, which spans six orders
of magnitude and could follow a power law but with an exponential cutoff.
\end{enumerate}
This being an article about power laws, I will not discuss further the
possible explanations for these distributions, but the scientist confronted
with a new set of data having a broad dynamic range and a highly skewed
distribution should certainly bear in mind that a power-law model is only
one of several possibilities for fitting it.

\section{The mathematics of power laws}
A continuous real variable with a power-law distribution has a probability
$p(x)\>\d x$ of taking a value in the interval from $x$ to $x+\d x$, where
\begin{equation}
p(x) = C x^{-\alpha},
\label{basicdef}
\end{equation}
with $\alpha>0$.  As we saw in Section~\ref{secexamples}, there must be
some lowest value $x_\mathrm{min}$ at which the power law is obeyed, and we
consider only the statistics of $x$ above this value.

\subsection{Normalization}
\label{normalization}
The constant~$C$ in Eq.~\eref{basicdef} is given by the normalization
requirement that
\begin{equation}
1 = \int_{x_\mathrm{min}}^\infty p(x)\>\d x
  = C \int_{x_\mathrm{min}}^\infty x^{-\alpha} \>\d x
  = {C\over1-\alpha} \Bigl[ x^{-\alpha+1} \Bigr]_{x_\mathrm{min}}^\infty.
\label{normcond}
\end{equation}
We see immediately that this only makes sense if $\alpha>1$, since
otherwise the right-hand side of the equation would diverge: power laws
with exponents less than unity cannot be normalized and don't normally
occur in nature.  If $\alpha>1$ then Eq.~\eref{normcond} gives
\begin{equation}
C = (\alpha-1) x_\mathrm{min}^{\alpha-1},
\label{normc}
\end{equation}
and the correct normalized expression for the power law itself is
\begin{equation}
p(x) = {\alpha-1\over x_\mathrm{min}}\,
        \biggl( {x\over x_\mathrm{min}} \biggr)^{-\alpha}.
\label{fullpx}
\end{equation}

Some distributions follow a power law for part of their range but are cut
off at high values of~$x$.  That is, above some value they deviate from the
power law and fall off quickly towards zero.  If this happens, then the
distribution may be normalizable no matter what the value of the
exponent~$\alpha$.  Even so, exponents less than unity are rarely, if ever,
seen.

\subsection{Moments}
\label{moments}
The mean value of our power-law distributed quantity~$x$ is given by
\begin{eqnarray}
\av{x} &=& \int_{x_\mathrm{min}}^\infty x p(x) \>\d x
        =  C \int_{x_\mathrm{min}}^\infty x^{-\alpha+1} \>\d x\nonumber\\
       &=& {C\over 2-\alpha} \Bigl[ x^{-\alpha+2}
                            \Bigr]_{x_\mathrm{min}}^\infty.
\label{firstmoment}
\end{eqnarray}
Note that this expression becomes infinite if $\alpha\le2$.  Power laws
with such low values of $\alpha$ have no finite mean.  The distributions of
sizes of solar flares and wars in Table~\ref{exponents} are examples of
such power laws.

What does it mean to say that a distribution has an infinite mean?  Surely
we can take the data for real solar flares and calculate their average?
Indeed we can and necessarily we will always get a finite number from the
calculation, since each individual measurement~$x$ is itself a finite
number and there are a finite number of them.  Only if we had a truly
infinite number of samples would we see the mean actually diverge.

However, if we were to repeat our finite experiment many times and
calculate the mean for each repetition, then the mean of those many means
is itself also formally divergent, since it is simply equal to the mean we
would calculate if all the repetitions were combined into one large
experiment.  This implies that, while the mean may take a relatively small
value on any particular repetition of the experiment, it must occasionally
take a huge value, in order that the overall mean diverge as the number of
repetitions does.  Thus there must be very large fluctuations in the value
of the mean, and this is what the divergence in Eq.~\eref{firstmoment}
really implies.  In effect, our calculations are telling us that the mean
is not a well defined quantity, because it can vary enormously from one
measurement to the next, and indeed can become arbitrarily large.  The
formal divergence of~$\langle x\rangle$ is a signal that, while we can
quote a figure for the average of the samples we measure, that figure is
not a reliable guide to the typical size of the samples in another instance
of the same experiment.

For $\alpha>2$ however, the mean is perfectly well defined, with a value
given by Eq.~\eref{firstmoment} of
\begin{equation}
\av{x} = {\alpha-1\over\alpha-2} x_\mathrm{min}.
\end{equation}

We can also calculate higher moments of the distribution~$p(x)$.  For
instance, the second moment, the mean square, is given by
\begin{equation}
\av{x^2} = {C\over 3-\alpha} \Bigl[ x^{-\alpha+3}
                            \Bigr]_{x_\mathrm{min}}^\infty.
\label{secondmoment}
\end{equation}
This diverges if $\alpha\le3$.  Thus power-law distributions in this range,
which includes almost all of those in Table~\ref{exponents}, have no
meaningful mean square, and thus also no meaningful variance or standard
deviation.  If $\alpha>3$, then the second moment is finite and
well-defined, taking the value
\begin{equation}
\av{x^2} = {\alpha-1\over\alpha-3} x_\mathrm{min}^2.
\end{equation}

These results can easily be extended to show that in general all moments
$\av{x^m}$ exist for $m<\alpha-1$ and all higher moments diverge.  The ones
that do exist are given by
\begin{equation}
\av{x^m} = {\alpha-1\over\alpha-1-m} x_\mathrm{min}^m.
\end{equation}

\subsection{Largest value}
\label{largest}
Suppose we draw $n$ measurements from a power-law distribution.  What value
is the largest of those measurements likely to take?  Or, more precisely,
what is the probability $\pi(x)\>\d x$ that the largest value falls in the
interval between $x$ and $x+\d x$?

The definitive property of the largest value in a sample is that there are
no others larger than it.  The probability that a particular sample will be
larger than $x$ is given by the quantity $P(x)$ defined in
Eq.~\eref{cumulative}:
\begin{equation}
P(x) = \int_x^\infty p(x') \>\d x' = {C\over\alpha-1} x^{-\alpha+1}
     = \biggl( {x\over x_\mathrm{min}} \biggr)^{-\alpha+1},
\label{expresspx}
\end{equation}
so long as $\alpha>1$.  And the probability that a sample is not greater
than $x$ is $1-P(x)$.  Thus the probability that a particular sample we
draw, sample~$i$, will lie between $x$ and $x+\d x$ and that all the others
will be no greater than it is $p(x)\>\d x\times[1-P(x)]^{n-1}$.  Then there are
$n$ ways to choose~$i$, giving a total probability
\begin{equation}
\pi(x) = n p(x) [1-P(x)]^{n-1}.
\end{equation}

Now we can calculate the mean value~$\av{x_\mathrm{max}}$ of the largest
sample thus:
\begin{equation}
\av{x_\mathrm{max}} = \int_{x_\mathrm{min}}^\infty x\pi(x) \>\d x
  = n \int_{x_\mathrm{min}}^\infty x p(x) [1 - P(x)]^{n-1} \>\d x.
\end{equation}
Using Eqs.~\eref{fullpx} and~\eref{expresspx}, this is
\begin{eqnarray}
\av{x_\mathrm{max}}
 &=& n(\alpha-1)\times\nonumber\\
 & & \hspace{0.4em} \int_{x_\mathrm{min}}^\infty 
     \biggl( {x\over x_\mathrm{min}} \biggr)^{-\alpha+1}
     \biggl[ 1 -  \biggl( {x\over x_\mathrm{min}} \biggr)^{-\alpha+1}
     \biggr]^{n-1} \d x \nonumber\\
 &=& nx_\mathrm{min} \int_0^1 {y^{n-1}\over(1-y)^{1/(\alpha-1)}} \>\d y
     \nonumber\\
 &=& nx_\mathrm{min}\,\Beta\bigl(n,(\alpha-2)/(\alpha-1)\bigr),
\label{xmax}
\end{eqnarray}
where I have made the substitution $y=1-(x/x_\mathrm{min})^{-\alpha+1}$ and
$\Beta(a,b)$ is Legendre's beta-function,\footnote{Also called the Eulerian
integral of the first kind.} which is defined by
\begin{equation}
\Beta(a,b) = {\Gamma(a)\Gamma(b)\over\Gamma(a+b)},
\label{defsbeta}
\end{equation}
with $\Gamma(a)$ the standard $\Gamma$-function:
\begin{equation}
\Gamma(a) = \int_0^\infty t^{a-1} \e^{-t} \>\d t.
\label{defsgamma}
\end{equation}

The beta-function has the interesting property that for large values of
either of its arguments it itself follows a power law.\footnote{This can be
demonstrated by approximating the $\Gamma$-functions of Eq.~\eref{defsbeta}
using Sterling's formula.}  For instance, for large $a$ and fixed~$b$,
$\Beta(a,b)\sim a^{-b}$.  In most cases of interest, the number~$n$ of
samples from our power-law distribution will be large (meaning much greater
than~1), so
\begin{equation}
\Beta\bigl(n,(\alpha-2)/(\alpha-1)\bigr) \sim n^{-(\alpha-2)/(\alpha-1)},
\end{equation}
and
\begin{equation}
\av{x_\mathrm{max}} \sim n^{1/(\alpha-1)}.
\label{asymptotic}
\end{equation}
Thus, as long as $\alpha>1$, we find that $\av{x_\mathrm{max}}$ always
increases as $n$ becomes larger.\footnote{Equation~\eref{asymptotic} can
also be derived by a simpler, although less rigorous, heuristic argument:
if $P(x)=1/n$ for some value of~$x$ then we expect there to be on average
one sample in the range from~$x$ to~$\infty$, and this of course will the
largest sample.  Thus a rough estimate of $\av{x_\mathrm{max}}$ can be
derived by setting our expression for $P(x)$, Eq.~\eref{expresspx}, equal
to~$1/n$ and rearranging for~$x$, which immediately gives
$\av{x_\mathrm{max}} \sim n^{1/(\alpha-1)}$.}

\subsection{Top-heavy distributions and the 80/20 rule}
Another interesting question is where the majority of the distribution of
$x$ lies.  For any power law with exponent $\alpha>1$, the median is
well defined.  That is, there is a point $x_{1/2}$ that divides the
distribution in half so that half the measured values of~$x$ lie above
$x_{1/2}$ and half lie below.  That point is given by
\begin{equation}
\int_{x_{1/2}}^\infty p(x)\>\d x = \half \int_{x_\mathrm{min}}^\infty
p(x)\>\d x,
\end{equation}
or
\begin{equation}
x_{1/2} = 2^{1/(\alpha-1)} x_\mathrm{min}.
\end{equation}

So, for example, if we are considering the distribution of wealth, there
will be some well-defined median wealth that divides the richer half of the
population from the poorer.  But we can also ask how much of the wealth
itself lies in those two halves.  Obviously more than half of the total
amount of money belongs to the richer half of the population.  The fraction
of the money in the richer half is given by
\begin{equation}
{\int_{x_{1/2}}^\infty xp(x)\>\d x\over
 \int_{x_\mathrm{min}}^\infty xp(x)\>\d x}
  = \biggl( {x_{1/2}\over x_\mathrm{min}} \biggr)^{-\alpha+2}
  = 2^{-(\alpha-2)/(\alpha-1)},
\label{richest}
\end{equation}
provided $\alpha>2$ so that the integrals converge.  Thus, for instance,
if $\alpha=2.1$ for the wealth distribution, as indicated in
Table~\ref{exponents}, then a fraction $2^{-0.091}\simeq94\%$ of the wealth
is in the hands of the richer 50\% of the population, making the
distribution quite top-heavy.

\begin{figure}
\begin{center}
\resizebox{7cm}{!}{\includegraphics{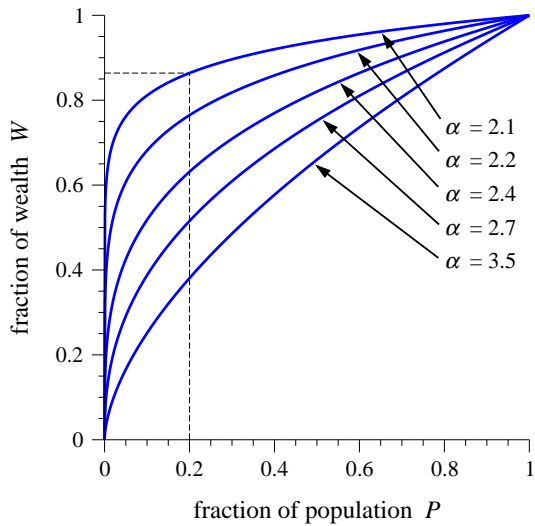}}
\end{center}
\caption{The fraction~$W$ of the total wealth in a country held by the
fraction~$P$ of the richest people, if wealth is distributed following a
power law with exponent~$\alpha$.  If $\alpha=2.1$, for instance, as it
appears to in the United States (Table~\ref{exponents}), then the richest
20\% of the population hold about 86\% of the wealth (dashed lines).}
\label{eighty20}
\end{figure}

More generally, the fraction of the population whose personal wealth
exceeds~$x$ is given by the quantity~$P(x)$, Eq.~\eref{expresspx}, and the
fraction of the \emph{total} wealth in the hands of those people is
\begin{equation}
W(x) = {\int_x^\infty x' p(x') \>\d x'\over
        \int_{x_\mathrm{min}}^\infty x' p(x') \>\d x'}
     = \biggl( {x\over x_\mathrm{min}} \biggr)^{-\alpha+2},
\label{defswx}
\end{equation}
assuming again that $\alpha>2$.  Eliminating $x/x_\mathrm{min}$
between~\eref{expresspx} and~\eref{defswx}, we find that the fraction~$W$
of the wealth in the hands of the richest~$P$ of the population is
\begin{equation}
W = P^{(\alpha-2)/(\alpha-1)},
\label{wpcurve}
\end{equation}
of which Eq.~\eref{richest} is a special case.  This again has a power-law
form, but with a positive exponent now.  In Fig.~\ref{eighty20} I show the
form of the curve of $W$ against $P$ for various values of~$\alpha$.  For
all values of~$\alpha$ the curve is concave downwards, and for values only
a little above~2 the curve has a very fast initial increase, meaning that a
large fraction of the wealth is concentrated in the hands of a small
fraction of the population.  Curves of this kind are called \defn{Lorenz
curves}, after Max Lorenz, who first studied them around the turn of the
twentieth century~\cite{Lorenz05}.

Using the exponents from Table~\ref{exponents}, we can for example
calculate that about 80\% of the wealth should be in the hands of the
richest 20\% of the population (the so-called ``80/20 rule'', which is
borne out by more detailed observations of the wealth distribution), the
top 20\% of web sites get about two-thirds of all web hits, and the largest
10\% of US cities house about 60\% of the country's total population.

If $\alpha\le2$ then the situation becomes even more extreme.  In that
case, the integrals in Eq.~\eref{defswx} diverge at their upper limits,
meaning that in fact they depend on the value of the largest sample, as
described in Section~\ref{moments}.  But for $\alpha>1$,
Eq.~\eref{asymptotic} tells us that the expected value of $x_\mathrm{max}$
goes to~$\infty$ as $n$ becomes large, and in that limit the fraction of
money in the top half of the population, Eq.~\eref{richest}, tends to
unity.  In fact, the fraction of money in the top \emph{anything} of the
population, even the top 1\%, tends to unity, as Eq.~\eref{defswx} shows.
In other words, for distributions with $\alpha<2$, essentially all of the
wealth (or other commodity) lies in the tail of the distribution.  The
distribution of family names in the US, which has an exponent~$\alpha=1.9$,
is an example of this type of behaviour.  For the data of
Fig.~\ref{examples}k, about 75\% of the population have names in the top
$15\,000$.  Estimates of the total number of unique family names in the US
put the figure at around 1.5 million.  So in this case 75\% of the
population have names in the most common~1\%---a very top-heavy
distribution indeed.  The line $\alpha=2$ thus separates the regime in
which you will with some frequency meet people with uncommon names from the
regime in which you will rarely meet such people.

\subsection{Scale-free distributions}
\label{secsf}
A power-law distribution is also sometimes called a \defn{scale-free
distribution}.  Why?  Because a power law is the only distribution that is
the same \emph{whatever scale we look at it on.}  By this we mean the
following.

Suppose we have some probability distribution~$p(x)$ for a quantity~$x$,
and suppose we discover or somehow deduce that it satisfies the property
that
\begin{equation}
p(bx) = g(b) p(x),
\label{scalefree}
\end{equation}
for any~$b$.  That is, if we increase the scale or units by which we
measure~$x$ by a factor of~$b$, the shape of the distribution~$p(x)$ is
unchanged, except for an overall multiplicative constant.  Thus for
instance, we might find that computer files of size 2kB are $\frac14$ as
common as files of size~1kB.  Switching to measuring size in megabytes we
also find that files of size 2MB are $\frac14$ as common as files of
size~1MB.  Thus the shape of the file-size distribution curve (at least for
these particular values) does not depend on the scale on which we measure
file size.

This scale-free property is certainly not true of most distributions.  It
is not true for instance of the exponential distribution.  In fact, as we
now show, it is only true of one type of distribution, the power law.

Starting from Eq.~\eref{scalefree}, let us first set $x=1$, giving
$p(b)=g(b) p(1)$.  Thus $g(b)=p(b)/p(1)$ and~\eref{scalefree} can be
written as
\begin{equation}
p(bx) = {p(b) p(x)\over p(1)}.
\end{equation}
Since this equation is supposed to be true for any~$b$, we can
differentiate both sides with respect to~$b$ to get
\begin{equation}
x p'(bx) = {p'(b) p(x)\over p(1)},
\end{equation}
where $p'$ indicates the derivative of $p$ with respect to its argument.
Now we set $b=1$ and get
\begin{equation}
x {\d p\over\d x} = {p'(1)\over p(1)} p(x).
\end{equation}
This is a simple first-order differential equation which has the solution
\begin{equation}
\ln p(x) = {p(1)\over p'(1)} \ln x + \mbox{constant}.
\end{equation}
Setting $x=1$ we find that the constant is simply $\ln p(1)$, and then
taking exponentials of both sides
\begin{equation}
p(x) = p(1)\,x^{-\alpha},
\end{equation}
where $\alpha = -p(1)/p'(1)$.  Thus, as advertised, the power-law
distribution is the only function satisfying the scale-free
criterion~\eref{scalefree}.

This fact is more than just a curiosity.  As we will see in
Section~\ref{critphen}, there are some systems that become scale-free for
certain special values of their governing parameters.  The point defined by
such a special value is called a ``continuous phase transition'' and the
argument given above implies that at such a point the observable quantities
in the system should adopt a power-law distribution.  This indeed is seen
experimentally and the distributions so generated provided the original
motivation for the study of power laws in physics (although most
experimentally observed power laws are probably not the result of phase
transitions---a variety of other mechanisms produce power-law behaviour as
well, as we will shortly see).

\subsection{Power laws for discrete variables}
So far I have focused on power-law distributions for continuous real
variables, but many of the quantities we deal with in practical situations
are in fact discrete---usually integers.  For instance, populations of
cities, numbers of citations to papers or numbers of copies of books sold
are all integer quantities.  In most cases, the distinction is not very
important.  The power law is obeyed only in the tail of the distribution
where the values measured are so large that, to all intents and purposes,
they can be considered continuous.  Technically however, power-law
distributions should be defined slightly differently for integer
quantities.

If $k$ is an integer variable, then one way to proceed is to declare that
it follows a power law if the probability~$p_k$ of measuring the value~$k$
obeys
\begin{equation}
p_k = C k^{-\alpha},
\label{discrete}
\end{equation}
for some constant exponent~$\alpha$.  Clearly this distribution cannot hold
all the way down to $k=0$, since it diverges there, but it could in theory
hold down to $k=1$.  If we discard any data for $k=0$, the constant~$C$
would then be given by the normalization condition
\begin{equation}
1 = \sum_{k=1}^\infty p_k = C \sum_{k=1}^\infty k^{-\alpha} =
C\zeta(\alpha),
\end{equation}
where $\zeta(\alpha)$ is the Riemann $\zeta$-function.  Rearranging, we
find that $C = 1/\zeta(\alpha)$ and
\begin{equation}
p_k = {k^{-\alpha}\over\zeta(\alpha)}.
\end{equation}
If, as is usually the case, the power-law behaviour is seen only in the
tail of the distribution, for values $k\ge k_\mathrm{min}$, then the
equivalent expression is
\begin{equation}
p_k = {k^{-\alpha}\over\zeta(\alpha,k_\mathrm{min})},
\end{equation}
where $\zeta(\alpha,k_\mathrm{min})=\sum_{k=k_\mathrm{min}}^\infty
k^{-\alpha}$ is the generalized or incomplete
$\zeta$-function.

Most of the results of the previous sections can be generalized to the case
of discrete variables, although the mathematics is usually harder and often
involves special functions in place of the more tractable integrals of the
continuous case.

It has occasionally been proposed that Eq.~\eref{discrete} is not the best
generalization of the power law to the discrete case.  An alternative and
often more convenient form is
\begin{equation}
p_k = C\,{\Gamma(k)\Gamma(\alpha)\over\Gamma(k+\alpha)}
    = C\,\Beta(k,\alpha),
\label{yuledist}
\end{equation}
where $\Beta(a,b)$ is, as before, the Legendre beta-function,
Eq.~\eref{defsbeta}.  As mentioned in Section~\ref{largest}, the
beta-function behaves as a power law $\Beta(k,\alpha) \sim k^{-\alpha}$ for
large~$k$ and so the distribution has the desired asymptotic form.
Simon~\cite{Simon55} proposed that Eq.~\eref{yuledist} be called the
\defn{Yule distribution}, after Udny Yule who derived it as the limiting
distribution in a certain stochastic process~\cite{Yule25}, and this name
is often used today.  Yule's result is described in Section~\ref{yule}.

The Yule distribution is nice because sums involving it can frequently be
performed in closed form, where sums involving Eq.~\eref{discrete} can only
be written in terms of special functions.  For instance, the normalizing
constant~$C$ for the Yule distribution is given by
\begin{equation}
1 = C \sum_{k=1}^\infty \Beta(k,\alpha) = {C\over\alpha-1},
\end{equation}
and hence $C=\alpha-1$ and
\begin{equation}
p_k = (\alpha-1)\,\Beta(k,\alpha).
\end{equation}
The first and second moments (i.e.,~the mean and mean square of the
distribution) are
\begin{equation}
\av{k} = {\alpha-1\over\alpha-2},\qquad
\av{k^2} = {(\alpha-1)^2\over(\alpha-2)(\alpha-3)},
\end{equation}
and there are similarly simple expressions corresponding to many of our
earlier results for the continuous case.

\section{Mechanisms for generating power-law distributions}
\label{generation}
In this section we look at possible candidate mechanisms by which power-law
distributions might arise in natural and man-made systems.  Some of the
possibilities that have been suggested are quite complex---notably the
physics of critical phenomena and the tools of the renormalization group
that are used to analyse it.  But let us start with some simple algebraic
methods of generating power-law functions and progress to the more involved
mechanisms later.

\subsection{Combinations of exponentials}
A much more common distribution than the power law is the exponential,
which arises in many circumstances, such as survival times for decaying
atomic nuclei or the Boltzmann distribution of energies in statistical
mechanics.  Suppose some quantity $y$ has an exponential distribution:
\begin{equation}
p(y) \sim \e^{ay}.
\label{exp1}
\end{equation}
The constant~$a$ might be either negative or positive.  If it is positive
then there must also be a cutoff on the distribution---a limit on the
maximum value of~$y$---so that the distribution is normalizable.

Now suppose that the real quantity we are interested in is not~$y$ but some
other quantity~$x$, which is exponentially related to $y$ thus:
\begin{equation}
x \sim \e^{by},
\label{exp2}
\end{equation}
with $b$ another constant, also either positive or negative.  Then the
probability distribution of $x$ is
\begin{equation}
p(x) = p(y) {\d y\over\d x} \sim {\e^{ay}\over b\e^{by}}
     = {x^{-1+a/b}\over b},
\label{expexp}
\end{equation}
which is a power law with exponent $\alpha=1-a/b$.

A version of this mechanism was used by Miller~\cite{Miller57} to explain
the power-law distribution of the frequencies of words as follows (see
also~\cite{Li92}).  Suppose we type randomly on a typewriter,\footnote{This
argument is sometimes called the ``monkeys with typewriters'' argument, the
monkey being the traditional exemplar of a random typist.} pressing the
space bar with probability~$q_s$ per stroke and each letter with equal
probability~$q_l$ per stroke.  If there are $m$ letters in the alphabet
then $q_l=(1-q_s)/m$.  (In this simplest version of the argument we also
type no punctuation, digits or other non-letter symbols.)  Then the
frequency~$x$ with which a particular word with $y$ letters (followed by a
space) occurs is
\begin{equation}
x = \biggl[ {1-q_s\over m} \biggr]^y q_s \sim \e^{by},
\end{equation}
where $b=\ln(1-q_s)-\ln m$.  The number (or fraction) of distinct possible
words with length between~$y$ and $y+\d y$ goes up exponentially as
$p(y)\sim m^y=\e^{ay}$ with $a=\ln m$.  Thus, following our argument above,
the distribution of frequencies of words has the form $p(x) \sim
x^{-\alpha}$ with
\begin{equation}
\alpha = 1-{a\over b} = {2\ln m - \ln(1-q_s)\over\ln m - \ln(1-q_s)}.
\end{equation}
For the typical case where $m$ is reasonably large and $q_s$ quite small
this gives $\alpha\simeq2$ in approximate agreement with
Table~\ref{exponents}.

\begin{figure}
\begin{center}
\resizebox{8cm}{!}{\includegraphics{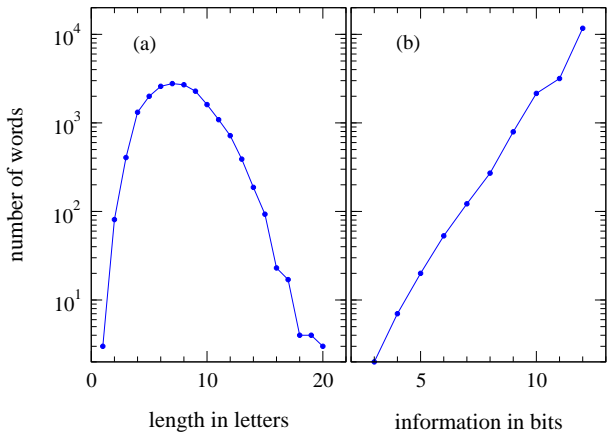}}
\end{center}
\caption{(a)~Histogram of the lengths in letters of all distinct words in
the text of the novel \textit{Moby Dick}.  (b)~Histogram of the information
content \textit{a la} Shannon of words in \textit{Moby Dick}.  The former
does not, by any stretch of the imagination, follow an exponential, but the
latter could easily be said to do so.  (Note that the vertical axes are
logarithmic.)}
\label{mdzipf}
\end{figure}

This is a reasonable theory as far as it goes, but real text is not made up
of random letters.  Most combinations of letters don't occur in natural
languages; most are not even pronounceable.  We might imagine that some
constant fraction of possible letter sequences of a given length would
correspond to real words and the argument above would then work just fine
when applied to that fraction, but upon reflection this suggestion is
obviously bogus.  It is clear for instance that very long words simply
don't exist in most languages, although there are exponentially many
possible combinations of letters available to make them up.  This
observation is backed up by empirical data.  In Fig.~\ref{mdzipf}a we show
a histogram of the lengths of words occurring in the text of \textit{Moby
Dick}, and one would need a particularly vivid imagination to convince
oneself that this histogram follows anything like the exponential assumed
by Miller's argument.  (In fact, the curve appears roughly to follow a
log-normal~\cite{LSA01}.)

There may still be some merit in Miller's argument however.  The problem
may be that we are measuring word ``length'' in the wrong units.  Letters
are not really the basic units of language.  Some basic units are letters,
but some are groups of letters.  The letters ``th'' for example often occur
together in English and make a single sound, so perhaps they should be
considered to be a separate symbol in their own right and contribute only
one unit to the word length?

Following this idea to its logical conclusion we can imagine replacing each
fundamental unit of the language---whatever that is---by its own symbol and
then measuring lengths in terms of numbers of symbols.  The pursuit of
ideas along these lines led Claude Shannon in the 1940s to develop the
field of information theory, which gives a precise prescription for
calculating the number of symbols necessary to transmit words or any other
data~\cite{Shannon48a,Shannon48b}.  The units of information are
\defn{bits} and the true ``length'' of a word can be considered to be the
number of bits of information it carries.  Shannon showed that if we regard
words as the basic divisions of a message, the information~$y$ carried by
any particular word is
\begin{equation}
y = -k\ln x,
\end{equation}
where $x$ is the frequency of the word as before and $k$ is a constant.
(The reader interested in finding out more about where this simple relation
comes from is recommended to look at the excellent introduction to
information theory by Cover and Thomas~\cite{CT91}.)

But this has precisely the form that we want.  Inverting it we have
$x=\e^{-y/k}$ and if the probability distribution of the ``lengths''
measured in terms of bits is also exponential as in Eq.~\eref{exp1} we will
get our power-law distribution.  Figure~\ref{mdzipf}b shows the latter
distribution, and indeed it follows a nice exponential---much better than
Fig.~\ref{mdzipf}a.

This is still not an entirely satisfactory explanation.  Having made the
shift from pure word length to information content, our simple count of the
\emph{number} of words of length~$y$---that it goes exponentially as
$m^y$---is no longer valid, and now we need some reason why there should be
exponentially more distinct words in the language of high information
content than of low.  That this is the case is experimentally verified by
Fig.~\ref{mdzipf}b, but the reason must be considered still a matter of
debate.  Some possibilities are discussed by, for instance,
Mandelbrot~\cite{Mandelbrot53} and more recently by
Mitzenmacher~\cite{Mitzenmacher04}.

Another example of the ``combination of exponentials'' mechanism has been
discussed by Reed and Hughes~\cite{RH02}.  They consider a process in
which a set of items, piles or groups each grows exponentially in time,
having size~$x\sim\e^{bt}$ with $b>0$.  For instance, populations of
organisms reproducing freely without resource constraints grow
exponentially.  Items also have some fixed probability of dying per unit
time (populations might have a stochastically constant probability of
extinction), so that the times $t$ at which they die are exponentially
distributed $p(t)\sim\e^{at}$ with $a<0$.

These functions again follow the form of Eqs.~\eref{exp1} and~\eref{exp2}
and result in a power-law distribution of the sizes~$x$ of the items or
groups at the time they die.  Reed and Hughes suggest that variations on
this argument may explain the sizes of biological taxa, incomes and cities,
among other things.

\subsection{Inverses of quantities}
Suppose some quantity $y$ has a distribution $p(y)$ that passes through
zero, thus having both positive and negative values.  And suppose further
that the quantity we are really interested in is the reciprocal $x=1/y$,
which will have distribution
\begin{equation}
p(x) = p(y) {\d y\over\d x} = - {p(y)\over x^2}.
\end{equation}
The large values of~$x$, those in the tail of the distribution, correspond
to the small values of $y$ close to zero and thus the large-$x$ tail is
given by
\begin{equation}
p(x) \sim x^{-2},
\end{equation}
where the constant of proportionality is~$p(y=0)$.

More generally, any quantity $x=y^{-\gamma}$ for some $\gamma$ will have a
power-law tail to its distribution $p(x) \sim x^{-\alpha}$, with
$\alpha=1+1/\gamma$.  It is not clear who the first author or authors were
to describe this mechanism,\footnote{A correspondent tells me that a
similar mechanism was described in an astrophysical context by
Chandrasekhar in a paper in 1943, but I have been unable to confirm this.}
but clear descriptions have been given recently by
Bouchaud~\cite{Bouchaud95}, Jan~\etal~\cite{JMRS99} and
Sornette~\cite{Sornette01}.

One might argue that this mechanism merely generates a power law by
assuming another one: the power-law relationship between $x$ and $y$
generates a power-law distribution for~$x$.  This is true, but the point is
that the mechanism takes some physical power-law relationship between $x$
and~$y$---\emph{not} a stochastic probability distribution---and from that
generates a power-law probability distribution.  This is a non-trivial
result.

One circumstance in which this mechanism arises is in measurements of the
fractional change in a quantity.  For instance, Jan~\etal~\cite{JMRS99}
consider one of the most famous systems in theoretical physics, the Ising
model of a magnet.  In its paramagnetic phase, the Ising model has a
magnetization that fluctuates around zero.  Suppose we measure the
magnetization~$m$ at uniform intervals and calculate the fractional change
$\delta=(\Delta m)/m$ between each successive pair of measurements.  The
change $\Delta m$ is roughly normally distributed and has a typical size
set by the width of that normal distribution.  The $1/m$ on the other hand
produces a power-law tail when small values of $m$ coincide with large
values of $\Delta m$, so that the tail of the distribution of $\delta$
follows $p(\delta)\sim\delta^{-2}$ as above.

\begin{figure}
\begin{center}
\resizebox{7cm}{!}{\includegraphics{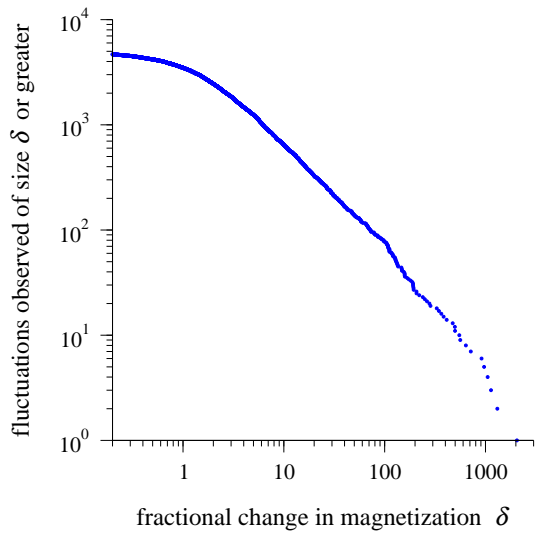}}
\end{center}
\caption{Cumulative histogram of the magnetization fluctuations of a
$128\times128$ nearest-neighbour Ising model on a square lattice.  The
model was simulated at a temperature of $2.5$ times the spin-spin coupling
for $100\,000$ time steps using the cluster algorithm of Swendsen and
Wang~\cite{SW87} and the magnetization per spin measured at intervals of
ten steps.  The fluctuations were calculated as the ratio
$\delta_i=2(m_{i+1}-m_i)/(m_{i+1}+m_i)$.}
\label{ising}
\end{figure}

In Fig.~\ref{ising} I show a cumulative histogram of measurements of
$\delta$ for simulations of the Ising model on a square lattice, and the
power-law distribution is clearly visible.  Using Eq.~\eref{mle}, the value
of the exponent is $\alpha=1.98\pm0.04$, in good agreement with the
expected value of~2.

\subsection{Random walks}
\label{rw}
Many properties of random walks are distributed according to power laws,
and this could explain some power-law distributions observed in nature.  In
particular, a randomly fluctuating process that undergoes ``gambler's
ruin'',\footnote{Gambler's ruin is so called because a gambler's night of
betting ends when his or her supply of money hits zero (assuming the
gambling establishment declines to offer him or her a line of credit).}
i.e.,~that ends when it hits zero, has a power-law distribution of possible
lifetimes.

\begin{figure}
\begin{center}
\resizebox{\columnwidth}{!}{\includegraphics{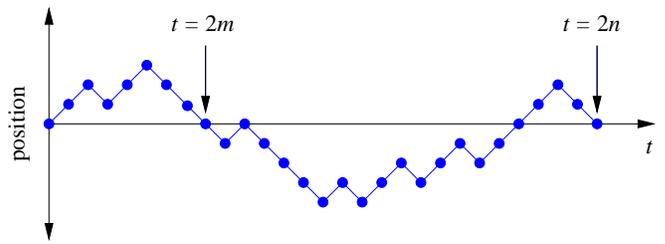}}
\end{center}
\caption{The position of a one-dimensional random walker (vertical axis) as
a function of time (horizontal axis).  The probability $u_{2n}$ that the
walk returns to zero at time~$t=2n$ is equal to the probability $f_{2m}$
that it returns to zero for the \emph{first time} at some earlier time
$t=2m$, multiplied by the probability $u_{2n-2m}$ that it returns again a
time $2n-2m$ later, summed over all possible values of~$m$.  We can use
this observation to write a consistency relation, Eq.~\eref{fpt}, that can
be solved for~$f_t$, Eq.~\eref{firstreturn}.}
\label{frt}
\end{figure}

Consider a random walk in one dimension, in which a walker takes a single
step randomly one way or the other along a line in each unit of time.
Suppose the walker starts at position~0 on the line and let us ask what the
probability is that the walker returns to position~0 for the first time at
time~$t$ (i.e.,~after exactly $t$ steps).  This is the so-called
\defn{first return time} of the walk and represents the lifetime of a
gambler's ruin process.  A trick for answering this question is depicted in
Fig.~\ref{frt}.  We consider first the unconstrained problem in which the
walk is allowed to return to zero as many times as it likes, before
returning there again at time~$t$.  Let us denote the probability of this
event as~$u_t$.  Let us also denote by $f_t$ the probability that the first
return time is~$t$.  We note that both of these probabilities are non-zero
only for even values of their arguments since there is no way to get back
to zero in any odd number of steps.

As Fig.~\ref{frt} illustrates, the probability $u_t=u_{2n}$, with $n$
integer, can be written
\begin{equation}
u_{2n} = \biggl\lbrace\begin{array}{ll}
           1                             & \qquad\mbox{if $n=0$,} \\
           \sum_{m=1}^n f_{2m} u_{2n-2m} & \qquad\mbox{if $n\ge1$,}
         \end{array}
\label{fpt}
\end{equation}
where $m$ is also an integer and we define $f_0=0$ and $u_0=1$.  This
equation can conveniently be solved for $f_{2n}$ using a generating
function approach.  We define
\begin{equation}
U(z) = \sum_{n=0}^\infty u_{2n} z^n,\quad
F(z) = \sum_{n=1}^\infty f_{2n} z^n.
\label{genfns}
\end{equation}
Then, multiplying Eq.~\eref{fpt} throughout by~$z^n$ and summing, we find
\begin{eqnarray}
U(z) &=& 1 + \sum_{n=1}^\infty \sum_{m=1}^n f_{2m} u_{2n-2m} z^n
         \nonumber\\
     &=& 1 + \sum_{m=1}^\infty f_{2m} z^m
                \sum_{n=m}^\infty u_{2n-2m} z^{n-m}\nonumber\\
     &=& 1 + F(z) U(z).
\end{eqnarray}
So
\begin{equation}
F(z) = 1-{1\over U(z)}.
\end{equation}

The function $U(z)$ however is quite easy to calculate.  The
probability~$u_{2n}$ that we are at position zero after $2n$ steps is
\begin{equation}
u_{2n} = 2^{-2n} {2n\choose n},
\end{equation}
so\footnote{The enthusiastic reader can easily derive this result for him
or herself by expanding $(1-z)^{-1/2}$ using the binomial theorem.}
\begin{equation}
U(z) = \sum_{n=0}^\infty {2n\choose n} {z^n\over4^n}
     = {1\over\sqrt{1-z}}.
\end{equation}
And hence
\begin{equation}
F(z) = {1-\sqrt{1-z}}.
\end{equation}
Expanding this function using the binomial theorem thus:
\begin{eqnarray}
F(z) &=& \half z + {\half\times\half\over2!} z^2
                 + {\half\times\half\times\frac32\over3!} z^3 + \ldots
         \nonumber\\
     &=& \sum_{n=1}^\infty {{2n\choose n}\over(2n-1)\,2^{2n}} z^n
\end{eqnarray}
and comparing this expression with Eq.~\eref{genfns}, we immediately see
that
\begin{equation}
f_{2n} = {{2n\choose n}\over(2n-1)\,2^{2n}},
\label{firstreturn}
\end{equation}
and we have our solution for the distribution of first return times.

Now consider the form of $f_{2n}$ for large~$n$.  Writing out the binomial
coefficient as ${2n\choose n} = (2n)!/(n!)^2$, we take logs thus:
\begin{equation}
\ln f_{2n} = \ln (2n)! - 2\ln n! - 2n\ln 2 - \ln(2n-1),
\end{equation}
and use Sterling's formula $\ln n!\simeq n\ln n - n + \half\ln n$ to get
$\ln f_{2n} \simeq \half\ln2 - \half\ln n - \ln(2n-1)$, or
\begin{equation}
f_{2n} \simeq \sqrt{2\over n(2n-1)^2}.
\end{equation}
In the limit $n\to\infty$, this implies that $f_{2n}\sim n^{-3/2}$, or
equivalently
\begin{equation}
f_t \sim t^{-3/2}.
\end{equation}
So the distribution of return times follows a power law with
exponent~$\alpha=\frac32$.  Note that the distribution has a divergent mean
(because $\alpha\le2$).  As discussed in Section~\ref{largest}, this
implies that the mean is finite for any finite sample but can take very
different values for different samples, so that the value measured for any
one sample gives little or no information about the value for any other.

As an example application, the random walk can be considered a simple model
for the lifetime of biological taxa.  A \defn{taxon} is a branch of the
evolutionary tree, a group of species all descended by repeated speciation
from a common ancestor.\footnote{Modern phylogenetic analysis, the
quantitative comparison of species' genetic material, can provide a picture
of the evolutionary tree and hence allow the accurate ``cladistic''
assignment of species to taxa.  For prehistoric species, however, whose
genetic material is not usually available, determination of evolutionary
ancestry is difficult, so classification into taxa is based instead on
morphology, i.e.,~on the shapes of organisms.  It is widely acknowledged
that such classifications are subjective and that the taxonomic assignments
of fossil species are probably riddled with errors.}  The ranks of the
Linnean hierarchy---genus, family, order and so forth---are examples of
taxa.  If a taxon gains and loses species at random over time, then the
number of species performs a random walk, the taxon becoming extinct when
the number of species reaches zero for the first (and only) time.  (This is
one example of ``gambler's ruin''.)  Thus the time for which taxa live
should have the same distribution as the first return times of random
walks.

In fact, it has been argued that the distribution of the lifetimes of
genera in the fossil record does indeed follow a power law~\cite{SBFJ95}.
The best fits to the available fossil data put the value of the exponent at
$\alpha=1.7\pm0.3$, which is in agreement with the simple random walk
model~\cite{NP03a}.\footnote{To be fair, I consider the power law for the
distribution of genus lifetimes to fall in the category of ``tenuous''
identifications to which I alluded in footnote~\ref{note1}.  This theory
should be taken with a pinch of salt.}

\subsection{The Yule process}
\label{yule}
One of the most convincing and widely applicable mechanisms for generating
power laws is the \defn{Yule process}, whose invention was, coincidentally,
also inspired by observations of the statistics of biological taxa as
discussed in the previous section.

In addition to having a (possibly) power-law distribution of lifetimes,
biological taxa also have a very convincing power-law distribution of
sizes.  That is, the distribution of the number of species in a genus,
family or other taxonomic group appears to follow a power law quite
closely.  This phenomenon was first reported by Willis and Yule in 1922 for
the example of flowering plants~\cite{WY22}.  Three years later,
Yule~\cite{Yule25} offered an explanation using a simple model that has
since found wide application in other areas.  He argued as follows.

Suppose first that new species appear but they never die; species are only
ever added to genera and never removed.  This differs from the random walk
model of the last section, and certainly from reality as well.  It is
believed that in practice all species and all genera become extinct in the
end.  But let us persevere; there is nonetheless much of worth in Yule's
simple model.

Species are added to genera by \defn{speciation}, the splitting of one
species into two, which is known to happen by a variety of mechanisms,
including competition for resources, spatial separation of breeding
populations and genetic drift.  If we assume that this happens at some
stochastically constant rate, then it follows that a genus with $k$ species
in it will gain new species at a rate proportional to~$k$, since each of
the $k$ species has the same chance per unit time of dividing in two.  Let
us further suppose that occasionally, say once every $m$ speciation events,
the new species produced is, by chance, sufficiently different from the
others in its genus as to be considered the founder member of an entire new
genus.  (To be clear, we define $m$~such that $m$ species are added to
pre-existing genera and then one species forms a new genus.  So $m+1$ new
species appear for each new genus and there are $m+1$ species per genus on
average.)  Thus the number of genera goes up steadily in this model, as
does the number of species within each genus.

We can analyse this Yule process mathematically as follows.\footnote{Yule's
analysis of the process was considerably more involved than the one
presented here, essentially because the theory of stochastic processes as
we now know it did not yet exist in his time.  The master equation method
we employ is a relatively modern innovation, introduced in this context by
Simon~\cite{Simon55}.}  Let us measure the passage of time in the model by
the number of genera~$n$.  At each time-step one new species founds a new
genus, thereby increasing $n$ by~1, and $m$ other species are added to
various pre-existing genera which are selected in proportion to the number
of species they already have.  We denote by $p_{k,n}$ the fraction of
genera that have $k$ species when the total number of genera is~$n$.  Thus
the \emph{number} of such genera is~$np_{k,n}$.  We now ask what the
probability is that the next species added to the system happens to be
added to a particular genus~$i$ having $k_i$ species in it already.  This
probability is proportional to~$k_i$, and so when properly normalized is
just $k_i/\sum_i k_i$.  But $\sum_i k_i$ is simply the total number of
species, which is $n(m+1)$.  Furthermore, between the appearance of the
$n$th and the $(n+1)$th genera, $m$~other new species are added, so the
probability that genus~$i$ gains a new species during this interval is
$mk_i/(n(m+1))$.  And the total expected number of genera of size~$k$ that
gain a new species in the same interval is
\begin{equation}
{mk\over n(m+1)} \times np_{k,n} = {m\over m+1} kp_{k,n}.
\end{equation}

Now we observe that the number of genera with $k$ species will decrease on
each time step by exactly this number, since by gaining a new species they
become genera with $k+1$ instead.  At the same time the number
\emph{increases} because of species that previously had $k-1$ species and
now have an extra one.  Thus we can write a \defn{master equation} for the
new number $(n+1)p_{k,n+1}$ of genera with $k$ species thus:
\begin{equation}
(n+1)p_{k,n+1} = np_{k,n} + {m\over m+1}
                 \bigl[ (k-1) p_{k-1,n} - k p_{k,n} \bigr].
\label{yule1}
\end{equation}
The only exception to this equation is for genera of size~1, which instead
obey the equation
\begin{equation}
(n+1)p_{1,n+1} = np_{1,n} + 1 - {m\over m+1} p_{1,n},
\label{yule2}
\end{equation}
since by definition exactly one new such genus appears on each time step.

Now we ask what form the distribution of the sizes of genera takes in the
limit of long times.  To do this we allow $n\to\infty$ and assume that the
distribution tends to some fixed value $p_k = \lim_{n\to\infty} p_{n,k}$
independent of~$n$.  Then Eq.~\eref{yule2} becomes $p_1 = 1 -
{mp_1/(m+1)}$, which has the solution
\begin{equation}
p_1 = {m+1\over2m+1}.
\label{p1}
\end{equation}
And Eq.~\eref{yule1} becomes
\begin{equation}
p_k = {m\over m+1} \bigl[ (k-1) p_{k-1} - k p_k \bigr],
\end{equation}
which can be rearranged to read
\begin{equation}
p_k = {k-1\over k+1+1/m}\,p_{k-1},
\end{equation}
and then iterated to get
\begin{eqnarray}
p_k &=& {(k-1)(k-2)\ldots1\over(k+1+1/m)(k+1/m)\ldots(3+1/m)}\,p_1\nonumber\\
    &=& (1+1/m) {(k-1)\ldots1\over(k+1+1/m)\ldots(2+1/m)},
\end{eqnarray}
where I have made use of Eq.~\eref{p1}.  This can be simplified further by
making use of a handy property of the $\Gamma$-function,
Eq.~\eref{defsgamma}, that $\Gamma(a)=(a-1)\Gamma(a-1)$.  Using this, and
noting that $\Gamma(1)=1$, we get
\begin{eqnarray}
p_k &=& (1+1/m) {\Gamma(k)\Gamma(2+1/m)\over\Gamma(k+2+1/m)} \nonumber\\
    &=& (1+1/m) \Beta(k,2+1/m),
\label{gammas}
\end{eqnarray}
where $\Beta(a,b)$ is again the beta-function, Eq.~\eref{defsbeta}.  This,
we note, is precisely the distribution defined in Eq.~\eref{yuledist},
which Simon called the Yule distribution.  Since the beta-function has a
power-law tail $\Beta(a,b)\sim a^{-b}$, we can immediately see that $p_k$
also has a power-law tail with an exponent
\begin{equation}
\alpha = 2 + {1\over m}.
\label{yuleexponent}
\end{equation}
The mean number $m+1$ of species per genus for the example of flowering
plants is about~3, making $m\simeq2$ and $\alpha\simeq2.5$.  The actual
exponent for the distribution found by Willis and Yule~\cite{WY22} is
$\alpha=2.5\pm0.1$, which is in excellent agreement with the theory.

Most likely this agreement is fortuitous, however.  The Yule process is
probably not a terribly realistic explanation for the distribution of the
sizes of genera, principally because it ignores the fact that species (and
genera) become extinct.  However, it has been adapted and generalized by
others to explain power laws in many other systems, most famously city
sizes~\cite{Simon55}, paper citations~\cite{Price76,KRL00}, and links to
pages on the world wide web~\cite{BA99b,DMS00}.  The most general form of
the Yule process is as follows.

Suppose we have a system composed of a collection of objects, such as
genera, cities, papers, web pages and so forth.  New objects appear every
once in a while as cities grow up or people publish new papers.  Each
object also has some property~$k$ associated with it, such as number of
species in a genus, people in a city or citations to a paper, that is
reputed to obey a power law, and it is this power law that we wish to
explain.  Newly appearing objects have some initial value of $k$ which we
will denote~$k_0$.  New genera initially have only a single species
$k_0=1$, but new towns or cities might have quite a large initial
population---a single person living in a house somewhere is unlikely to
constitute a town in their own right but $k_0=100$ people might do so.  The
value of $k_0$ can also be zero in some cases: newly published papers
usually have zero citations for instance.

In between the appearance of one object and the next, $m$~new
species/people/citations etc.\ are added to the entire system.  That is
some cities or papers will get new people or citations, but not necessarily
all will.  And in the simplest case these are added to objects in
proportion to the number that the object already has.  Thus the probability
of a city gaining a new member is proportional to the number already there;
the probability of a paper getting a new citation is proportional to the
number it already has.  In many cases this seems like a natural process.
For example, a paper that already has many citations is more likely to be
discovered during a literature search and hence more likely to be cited
again.  Simon~\cite{Simon55} dubbed this type of ``rich-get-richer''
process the \defn{Gibrat principle}.  Elsewhere it also goes by the names
of the \defn{Matthew effect}~\cite{Merton68}, \defn{cumulative
advantage}~\cite{Price76}, or \defn{preferential attachment}~\cite{BA99b}.

There is a problem however when $k_0=0$.  For example, if new papers appear
with no citations and garner citations in proportion to the number they
currently have, which is zero, then no paper will ever get any citations!
To overcome this problem one typically assigns new citations not in
proportion simply to~$k$, but to $k+c$, where $c$ is some constant.  Thus
there are three parameters $k_0$, $c$ and~$m$ that control the behaviour of
the model.
\begin{widetext}
By an argument exactly analogous to the one given above, one can then
derive the master equation
\begin{equation}
(n+1) p_{k,n+1} = n p_{k,n} + m {k-1+c\over k_0+c+m} p_{k-1,n}
                            - m {k+c\over k_0+c+m} p_{k,n},
                  \qquad\mbox{for $k>k_0$,}
\label{yulegen1}
\end{equation}
and
\begin{equation}
(n+1) p_{k_0,n+1} = n p_{k_0,n} + 1 - m {k_0+c\over k_0+c+m} p_{k_0,n},
                    \qquad\mbox{for $k=k_0$.}
\label{yulegen2}
\end{equation}
(Note that $k$ is never less than~$k_0$, since each object appears with
$k=k_0$ initially.)

\end{widetext}

Looking for stationary solutions of these equations as before, we define
$p_k = \lim_{n\to\infty} p_{n,k}$ and find that
\begin{equation}
p_{k_0} = {k_0+c+m\over(m+1)(k_0+c)+m},
\label{genp0}
\end{equation}
and
\begin{eqnarray}
p_k &=& {(k-1+c)(k-2+c)\ldots(k_0+c)\over
         (k-1+c+\alpha)(k-2+c+\alpha)\ldots(k_0+c+\alpha)}\,p_{k_0} \nonumber\\
    &=& {\Gamma(k+c)\Gamma(k_0+c+\alpha)\over
         \Gamma(k_0+c)\Gamma(k+c+\alpha)}\,p_{k_0},
\end{eqnarray}
where I have made use of the $\Gamma$-function notation introduced for
Eq.~\eref{gammas} and, for reasons that will become clear in just moment, I
have defined $\alpha=2+(k_0+c)/m$.  As before, this expression can also be
written in terms of the beta-function, Eq.\eref{defsbeta}:
\begin{equation}
p_k = {\Beta(k+c,\alpha)\over\Beta(k_0+c,\alpha)}\,p_{k_0}.
\end{equation}
Since the beta-function follows a power law in its tail, $\Beta(a,b)\sim
a^{-b}$, the general Yule process generates a power-law distribution
$p_k\sim k^{-\alpha}$ with exponent related to the three parameters of the
process according to
\begin{equation}
\alpha = 2 + {k_0+c\over m}.
\end{equation}

For example, the original Yule process for number of species per genus has
$c=0$ and $k_0=1$, which reproduces the result of Eq.~\eref{yuleexponent}.
For citations of papers or links to web pages we have $k_0=0$ and we must
have $c>0$ to get any citations or links at all.  So $\alpha=2+c/m$.  In
his work on citations Price~\cite{Price76} assumed that $c=1$, so that
paper citations have the same exponent $\alpha=2+1/m$ as the standard Yule
process, although there doesn't seem to be any very good reason for making
this assumption.  As we saw in Table~\ref{exponents} (and as Price himself
also reported), real citations seem to have an exponent $\alpha\simeq3$, so
we should expect $c\simeq m$.  For the data from the Science Citation Index
examined in Section~\ref{secexamples}, the mean number~$m$ of citations per
paper is $8.6$.  So we should put $c\simeq8.6$ too if we want the Yule
process to match the observed exponent.

The most widely studied model of links on the web, that of Barab\'asi and
Albert~\cite{BA99b}, assumes $c=m$ so that $\alpha=3$, but again there
doesn't seem to be a good reason for this assumption.  The measured
exponent for numbers of links to web sites is about $\alpha=2.2$, so if the
Yule process is to match the data in this case, we should put
$c\simeq0.2m$.

However, the important point is that the Yule process is a plausible and
general mechanism that can explain a number of the power-law distributions
observed in nature and can produce a wide range of exponents to match the
observations by suitable adjustments of the parameters.  For several of the
distributions shown in Fig.~\ref{examples}, especially citations, city
populations and personal income, it is now the most widely accepted theory.

\subsection{Phase transitions and critical phenomena}
\label{critphen}
A completely different mechanism for generating power laws, one that has
received a huge amount of attention over the past few decades from the
physics community, is that of critical phenomena.

Some systems have only a single macroscopic length-scale, size-scale or
time-scale governing them.  A classic example is a magnet, which has a
\defn{correlation length} that measures the typical size of magnetic
domains.  Under certain circumstances this length-scale can diverge,
leaving the system with no scale at all.  As we will now see, such a system
is ``scale-free'' in the sense of Section~\ref{secsf} and hence the
distributions of macroscopic physical quantities have to follow power laws.
Usually the circumstances under which the divergence takes place are very
specific ones.  The parameters of the system have to be tuned very
precisely to produce the power-law behaviour.  This is something of a
disadvantage; it makes the divergence of length-scales an unlikely
explanation for generic power-law distributions of the type highlighted in
this paper.  As we will shortly see, however, there are some elegant and
interesting ways around this problem.

The precise point at which the length-scale in a system diverges is called
a \defn{critical point} or a \defn{phase transition}.  More specifically it
is a \defn{continuous} phase transition.  (There are other kinds of phase
transitions too.)  Things that happen in the vicinity of continuous phase
transitions are known as \defn{critical phenomena}, of which power-law
distributions are one example.

\begin{figure}
\begin{center}
\resizebox{5cm}{!}{\includegraphics{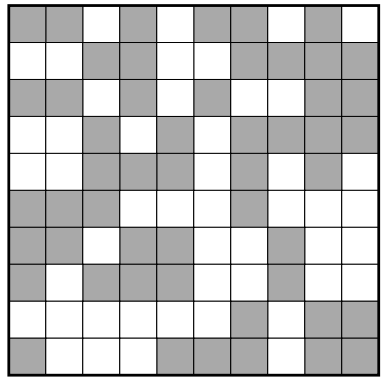}}
\end{center}
\caption{The percolation model on a square lattice: squares on the lattice
are coloured in independently at random with some probability~$p$.  In this
example $p=\half$.}
\label{percsmall}
\end{figure}

To better understand the physics of critical phenomena, let us explore one
simple but instructive example, that of the ``percolation transition''.
Consider a square lattice like the one depicted in Fig.~\ref{percsmall} in
which some of the squares have been coloured in.  Suppose we colour each
square with independent probability~$p$, so that on average a fraction~$p$
of them are coloured in.  Now we look at the \defn{clusters} of coloured
squares that form, i.e.,~the contiguous regions of adjacent coloured
squares.  We can ask, for instance, what the mean area~$\av{s}$ is of the
cluster to which a randomly chosen square belongs.  If that square is not
coloured in then the area is zero.  If it is coloured in but none of the
adjacent ones is coloured in then the area is one, and so forth.

\begin{figure*}
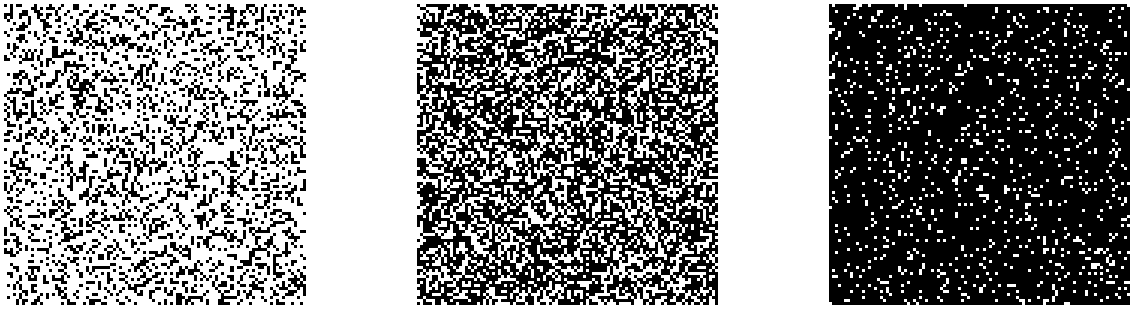

\begin{center}
\hfill\resizebox{4cm}{!}{\includegraphics{perc0.3.eps}}\hfill
\resizebox{4cm}{!}{\includegraphics{percpc.eps}}\hfill
\resizebox{4cm}{!}{\includegraphics{perc0.9.eps}}\hfill\null
\end{center}
\caption{Three examples of percolation systems on $100\times100$ square
lattices with $p=0.3$, $p=p_c=0.5927\ldots$ and $p=0.9$.  The first and
last are well below and above the critical point respectively, while the
middle example is precisely at it.}
\label{percolation}
\end{figure*}

When $p$ is small, only a few squares are coloured in and most coloured
squares will be alone on the lattice, or maybe grouped in twos or threes.
So $\av{s}$ will be small.  This situation is depicted in
Fig.~\ref{percolation} for $p=0.3$.  Conversely, if $p$ is
large---almost~1, which is the largest value it can have---then most
squares will be coloured in and they will almost all be connected together
in one large cluster, the so-called \defn{spanning cluster}.  In this
situation we say that the system \defn{percolates}.  Now the mean size of
the cluster to which a vertex belongs is limited only by the size of the
lattice itself and as we let the lattice size become large $\av{s}$ also
becomes large.  So we have two distinctly different behaviours, one for
small~$p$ in which $\av{s}$ is small and doesn't depend on the size of the
system, and one for large~$p$ in which $\av{s}$ is much larger and
increases with the size of the system.

\begin{figure}[b]
\begin{center}
\resizebox{8cm}{!}{\includegraphics{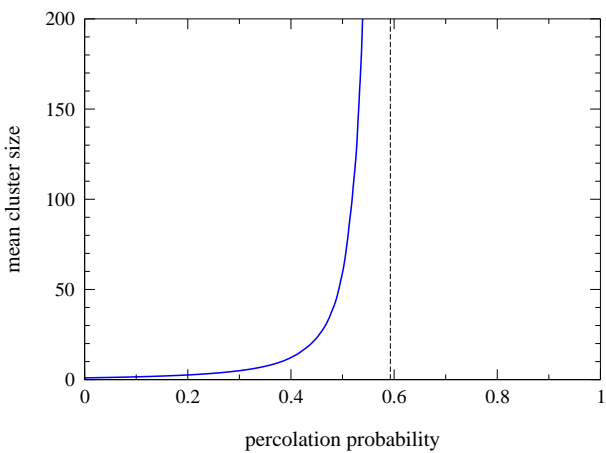}}
\end{center}
\caption{The mean area of the cluster to which a randomly chosen square
belongs for the percolation model described in the text, calculated from an
average over 1000 simulations on a $1000\times1000$ square lattice.  The
dotted line marks the known position of the phase transition.}
\label{divergence}
\end{figure}

And what happens in between these two extremes?  As we increase $p$ from
small values, the value of $\av{s}$ also increases.  But at some point we
reach the start of the regime in which $\av{s}$ goes up with system size
instead of staying constant.  We now know that this point is at
$p=0.5927462\ldots$, which is called the \defn{critical value} of~$p$ and
is denoted~$p_c$.  If the size of the lattice is large, then $\av{s}$ also
becomes large at this point, and in the limit where the lattice size goes
to infinity $\av{s}$~actually diverges.  To illustrate this phenomenon, I
show in Fig.~\ref{divergence} a plot of $\av{s}$ from simulations of the
percolation model and the divergence is clear.

Now consider not just the mean cluster size but the entire distribution of
cluster sizes.  Let $p(s)$ be the probability that a randomly chosen square
belongs to a cluster of area~$s$.  In general, what forms can $p(s)$ take
as a function of~$s$?  The important point to notice is that~$p(s)$, being
a probability distribution, is a dimensionless quantity---just a
number---but $s$ is an area.  We could measure~$s$ in terms of square
metres, or whatever units the lattice is calibrated in.  The
average~$\av{s}$ is also an area and then there is the area of a unit
square itself, which we will denote~$a$.  Other than these three
quantities, however, there are no other independent parameters with
dimensions in this problem.  (There is the area of the whole lattice, but
we are considering the limit where that becomes infinite, so it's out of
the picture.)

If we want to make a dimensionless function $p(s)$ out of these three
dimensionful parameters, there are three dimensionless ratios we can form:
$s/a$, $a/\av{s}$ and $s/\av{s}$ (or their reciprocals, if we prefer).
Only two of these are independent however, since the last is the product of
the other two.  Thus in general we can write
\begin{equation}
p(s) = C f\biggl( {s\over a}, {a\over\av{s}} \biggr),
\label{scaling}
\end{equation}
where $f$ is a dimensionless mathematical function of its dimensionless
arguments and $C$ is a normalizing constant chosen so that $\sum_s p(s)=1$.

\begin{figure}[b]
\begin{center}
\resizebox{\columnwidth}{!}{\includegraphics{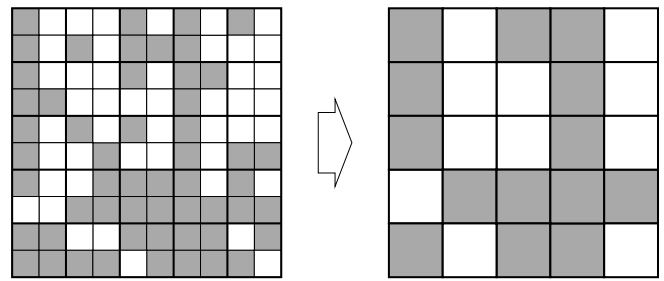}}
\end{center}
\caption{A site percolation system is coarse-grained, so that the area of
the fundamental square is (in this case) quadrupled.  The occupation of the
squares in the coarse-grained lattice (right) is chosen to mirror as nearly
as possible that of the squares on the original lattice (left), so that the
sizes and shapes of the large clusters remain roughly the same.  The small
clusters are mostly lost in the coarse-graining, so that the arguments
given in the text are valid only for the large-$s$ tail of the cluster size
distribution.}
\label{rescale}
\end{figure}

But now here's the trick.  We can \defn{coarse-grain} or \defn{rescale} our
lattice so that the fundamental unit of the lattice changes.  For instance,
we could double the size of our unit square~$a$.  The kind of picture I'm
thinking of is shown in Fig.~\ref{rescale}.  The basic percolation clusters
stay roughly the same size and shape, although I've had to fudge things
around the edges a bit to make it work.  For this reason this argument will
only be strictly correct for large clusters~$s$ whose area is not changed
appreciably by the fudging.  (And the argument thus only tells us that the
tail of the distribution is a power law, and not the whole distribution.)

The probability~$p(s)$ of getting a cluster of area~$s$ is unchanged by the
coarse-graining since the areas themselves are, to a good approximation,
unchanged, and the mean cluster size is thus also unchanged.  All that has
changed, mathematically speaking, is that the unit area $a$ has been
rescaled $a\to a/b$ for some constant rescaling factor~$b$.  The equivalent
of Eq.~\eref{scaling} in our coarse-grained system is
\begin{equation}
p(s) = C' f\biggl( {s\over a/b}, {a/b\over\av{s}} \biggr)
     = C' f\biggl( {bs\over a}, {a\over b\av{s}} \biggr).
\label{eqrescale}
\end{equation}
Comparing with Eq.~\eref{scaling}, we can see that this is equal, to within
a multiplicative constant, to the probability $p(bs)$ of getting a cluster
of size~$bs$, but in a system with a different mean cluster size of
$b\av{s}$.  Thus we have related the probabilities of two different sizes
of clusters to one another, but on systems with different average cluster
size and hence presumably also different site occupation probability.  Note
that the normalization constant must in general be changed in
Eq.~\eref{eqrescale} to make sure that $p(s)$ still sums to unity, and that
this change will depend on the value we choose for the rescaling
factor~$b$.

But now we notice that there is one special point at which this rescaling
by definition does not result in a change in $\av{s}$ or a corresponding
change in the site occupation probability, and that is the critical point.
When we are precisely at the point at which $\av{s}\to\infty$, then
$b\av{s}=\av{s}$ by definition.  Putting $\av{s}\to\infty$ in
Eqs.~\eref{scaling} and~\eref{eqrescale}, we then get $p(s) = C'
f({bs/a},0) = (C'/C) p(bs)$.  Or equivalently
\begin{equation}
p(bs) = g(b) p(s),
\end{equation}
where $g(b)=C/C'$.  Comparing with Eq.~\eref{scalefree} we see that this
has precisely the form of the equation that defines a scale-free
distribution.  The rest of the derivation below Eq.~\eref{scalefree}
follows immediately, and so we know that $p(s)$ must follow a power law.

This in fact is the origin of the name ``scale-free'' for a distribution of
the form~\eref{scalefree}.  At the point at which $\av{s}$ diverges, the
system is left with no defining size-scale, other than the unit of area~$a$
itself.  It is ``scale-free'', and by the argument above it follows that
the distribution of $s$ must obey a power law.

\begin{figure}[t]
\begin{center}
\resizebox{7cm}{!}{\includegraphics{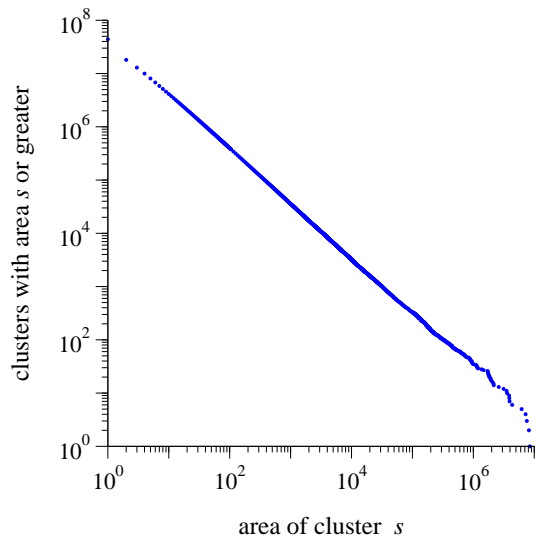}}
\end{center}
\caption{Cumulative distribution of the sizes of clusters for (site)
percolation on a square lattice of $40\,000\times40\,000$ sites at the
critical site occupation probability $p_c=0.592746\ldots$}
\label{pcdist}
\end{figure}

In Fig.~\ref{pcdist} I show an example of a cumulative distribution of
cluster sizes for a percolation system right at the critical point and, as
the figure shows, the distribution does indeed follow a power law.
Technically the distribution cannot follow a power law to arbitrarily large
cluster sizes since the area of a cluster can be no bigger than the area of
the whole lattice, so the power-law distribution will be cut off in the
tail.  This is an example of a \defn{finite-size effect}.  This point does
not seem to be visible in Fig.~\ref{pcdist} however.

The kinds of arguments given in this section can be made more precise using
the machinery of the \defn{renormalization group}.  The \defn{real-space
renormalization group} makes use precisely of transformations such as that
shown in Fig.~\ref{rescale} to derive power-law forms and their exponents
for distributions at the critical point.  An example application to the
percolation problem is given by Reynolds~\etal~\cite{RKS77}.  A more
technically sophisticated technique is the \defn{$k$-space renormalization
group}, which makes use of transformations in Fourier space to accomplish
similar aims in a particularly elegant formal environment~\cite{WK74}.

\subsection{Self-organized criticality}
\label{soc}
As discussed in the preceding section, certain systems develop power-law
distributions at special ``critical'' points in their parameter space
because of the divergence of some characteristic scale, such as the mean
cluster size in the percolation model.  This does not, however, provide a
plausible explanation for the origin of power laws in most real systems.
Even if we could come up with some model of earthquakes or solar flares or
web hits that had such a divergence, it seems unlikely that the parameters
of the real world would, just coincidentally, fall precisely at the point
where the divergence occurred.

As first proposed by Bak~\etal~\cite{BTW87}, however, it is possible that
some dynamical systems actually arrange themselves so that they always sit
at the critical point, no matter what state we start off in.  One says that
such systems \defn{self-organize} to the critical point, or that they
display \defn{self-organized criticality}.  A now-classic example of such a
system is the \defn{forest fire model} of Drossel and Schwabl~\cite{DS92},
which is based on the percolation model we have already seen.

\begin{figure}
\begin{center}
\resizebox{\columnwidth}{!}{\includegraphics{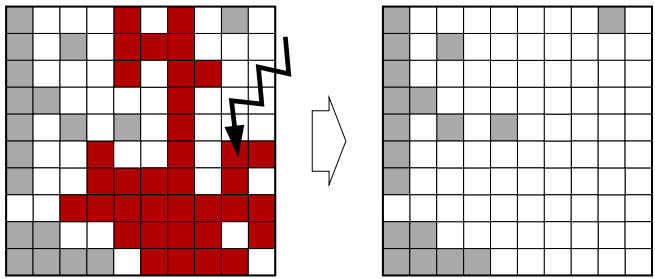}}
\end{center}
\caption{Lightning strikes at random positions in the forest fire model,
starting fires that wipe out the entire cluster to which a struck tree
belongs.}
\label{ffmodel}
\end{figure}

Consider the percolation model as a primitive model of a forest.  The
lattice represents the landscape and a single tree can grow in each square.
Occupied squares represent trees and empty squares represent empty plots of
land with no trees.  Trees appear instantaneously at random at some
constant rate and hence the squares of the lattice fill up at random.
Every once in a while a wildfire starts at a random square on the lattice,
set off by a lightning strike perhaps, and burns the tree in that square,
if there is one, along with every other tree in the cluster connected to
it.  The process is illustrated in Fig.~\ref{ffmodel}.  One can think of
the fire as leaping from tree to adjacent tree until the whole cluster is
burned, but the fire cannot cross the firebreak formed by an empty square.
If there is no tree in the square struck by the lightning, then nothing
happens.  After a fire, trees can grow up again in the squares vacated by
burnt trees, so the process keeps going indefinitely.

If we start with an empty lattice, trees will start to appear but will
initially be sparse and lightning strikes will either hit empty squares or
if they do chance upon a tree they will burn it and its cluster, but that
cluster will be small and localized because we are well below the
percolation threshold.  Thus fires will have essentially no effect on the
forest.  As time goes by however, more and more trees will grow up until at
some point there are enough that we have percolation.  At that point, as we
have seen, a spanning cluster forms whose size is limited only by the size
of the lattice, and when any tree in that cluster gets hit by the lightning
the entire cluster will burn away.  This gets rid of the spanning cluster
so that the system does not percolate any more, but over time as more trees
appear it will presumably reach percolation again, and so the scenario will
play out repeatedly.  The end result is that the system oscillates right
around the critical point, first going just above the percolation threshold
as trees appear and then being beaten back below it by fire.  In the limit
of large system size these fluctuations become small compared to the size
of the system as a whole and to an excellent approximation the system just
sits at the threshold indefinitely.  Thus, if we wait long enough, we
expect the forest fire model to self-organize to a state in which it has a
power-law distribution of the sizes of clusters, or of the sizes of fires.

\begin{figure}
\begin{center}
\resizebox{7cm}{!}{\includegraphics{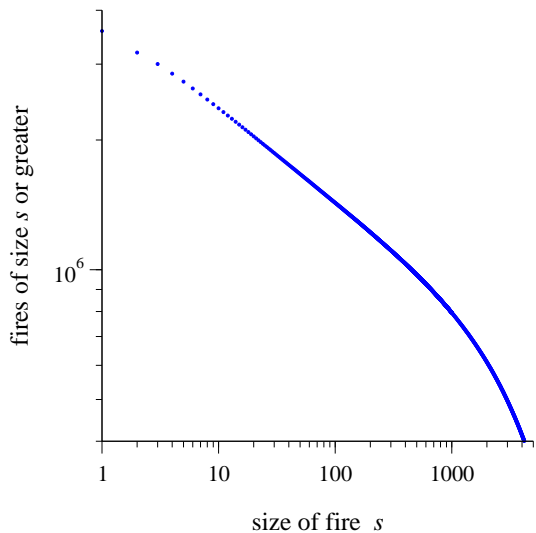}}
\end{center}
\caption{Cumulative distribution of the sizes of ``fires'' in a simulation
of the forest fire model of Drossel and Schwabl~\cite{DS92} for a square
lattice of size $5000\times5000$.}
\label{ffdist}
\end{figure}

In Fig.~\ref{ffdist} I show the cumulative distribution of the sizes of
fires in the forest fire model and, as we can see, it follows a power law
closely.  The exponent of the distribution is quite small in this case.
The best current estimates give a value of
$\alpha=1.19\pm0.01$~\cite{Grassberger02b}, meaning that the distribution
has an infinite mean in the limit of large system size.  For all real
systems however the mean is finite: the distribution is cut off in the
large-size tail because fires cannot have a size any greater than that of
the lattice as a whole and this makes the mean well-behaved.  This cutoff
is clearly visible in Fig.~\ref{ffdist} as the drop in the curve towards
the right of the plot.  What's more the distribution of the sizes of fires
in real forests, Fig.~\ref{nonpl}d, shows a similar cutoff and is in many
ways qualitatively similar to the distribution predicted by the model.
(Real forests are obviously vastly more complex than the forest fire model,
and no one is seriously suggesting that the model is an accurate
representation the real world.  Rather it is a guide to the general type of
processes that might be going on in forests.)

There has been much excitement about self-organized criticality as a
possible generic mechanism for explaining where power-law distributions
come from.  Per Bak, one of the originators of the idea, wrote an entire
book about it~\cite{Bak96}.  Self-organized critical models have been put
forward not only for forest fires, but for earthquakes~\cite{BT89,OFC92},
solar flares~\cite{LH91}, biological evolution~\cite{BS93},
avalanches~\cite{BTW87} and many other phenomena.  Although it is probably
not the universal law that some have claimed it to be, it is certainly a
powerful and intriguing concept that potentially has applications to a
variety of natural and man-made systems.

\subsection{Other mechanisms for generating power laws}
\label{sundry}
In the preceding sections I've described the best known and most widely
applied mechanisms that generate power-law distributions.  However, there
are a number of others that deserve a mention.  One that has been receiving
some attention recently is the \defn{highly optimized tolerance} mechanism
of Carlson and Doyle~\cite{CD99,CD00}.  The classic example of this
mechanism is again a model of forest fires and is based on the percolation
process.  Suppose again that fires start at random in a grid-like forest,
just as we considered in Sec.~\ref{soc}, but suppose now that instead of
appearing at random, trees are deliberately planted by a knowledgeable
forester.  One can ask what the best distribution of trees is to optimize
the amount of lumber the forest produces, subject to random fires that
could start at any place.  The answer turns out to be that one should plant
trees in blocks, with narrow firebreaks between them to prevent fires from
spreading.  Moreover, one should make the blocks smaller in regions where
fires start more often and larger where fires are rare.  The reason for
this is that we waste some valuable space by making firebreaks, space in
which we could have planted more trees.  If fires are rare, then on average
it pays to put the breaks further apart---more trees will burn if there is
a fire, but we also get more lumber if there isn't.

Carlson and Doyle show both by analytic arguments and by numerical
simulation that for quite general distributions of starting points for
fires this process leads to a distribution of fire sizes that approximately
follows a power law.  The distribution is not a perfect power law in this
case, but on the other hand neither are many of those seen in the data of
Fig.~\ref{examples}, so this is not necessarily a disadvantage.  Carlson
and Doyle have proposed that highly optimized tolerance could be a model
not only for forest fires but also for the sizes of files on the world wide
web, which appear to follow a power law~\cite{CB96}.

Another mechanism, which is mathematically similar to that of Carlson and
Doyle but quite different in motivation, is the \defn{coherent noise}
mechanism proposed by Sneppen and Newman~\cite{SN97} as a model of
biological extinction.  In this mechanism a number of agents or species are
subjected to stresses of various sizes, and each agent has a threshold for
stress above which an applied stress will wipe that agent out---the species
becomes extinct.  Extinct species are replaced by new ones with randomly
chosen thresholds.  The net result is that the system self-organizes to a
state where most of the surviving species have high thresholds, but the
exact distribution depends on the distribution of stresses in a way very
similar to the relation between block sizes and fire frequency in highly
optimized tolerance.  No conscious optimization is needed in this case, but
the end result is similar: the overall distribution of the numbers of
species becoming extinct as a result of any particular stress approximately
follows a power law.  The power-law form is not exact, but it's as good as
that seen in real extinction data.  Sneppen and Newman have also suggested
that their mechanism could be used to model avalanches and earthquakes.

One of the broad distributions mentioned in Sec.~\ref{secnonpl} as an
alternative to the power law was the log-normal.  A log-normally
distributed quantity is one whose logarithm is normally distributed.  That
is
\begin{equation}
p(\ln x) \sim \exp\biggl( - {(\ln x - \mu)^2\over2\sigma^2} \biggr),
\label{lognormal}
\end{equation}
for some choice of the mean~$\mu$ and standard deviation~$\sigma$ of the
distribution.  Distributions like this typically arise when we are
multiplying together random numbers.  The log of the product of a large
number of random numbers is the sum of the logarithms of those same random
numbers, and by the central limit theorem such sums have a normal
distribution essentially regardless of the distribution of the individual
numbers.

But Eq.~\eref{lognormal} implies that the distribution of~$x$ itself
is
\begin{equation}
p(x) = p(\ln x) {\d\ln x\over\d x}
     = {1\over x} \exp\biggl(-{(\ln x-\mu)^2\over2\sigma^2}\biggr).
\end{equation}
To see how this looks if we were to plot it on log scales, we take
logarithms of both sides, giving
\begin{eqnarray}
\ln p(x) &=& -\ln x - {(\ln x-\mu)^2\over2\sigma^2} \nonumber\\
         &=& - {(\ln x)^2\over2\sigma^2}
             + \biggl[ {\mu\over\sigma^2} - 1 \biggr] \ln x
             - {\mu^2\over2\sigma^2},
\label{lnquad}
\end{eqnarray}
which is quadratic in~$\ln x$.  However, any quadratic curve looks straight
if we view a sufficient small portion of it, so $p(x)$ will look like a
power-law distribution when we look at a small portion on log scales.  The
effective exponent~$\alpha$ of the distribution is in this case not fixed
by the theory---it could be anything, depending on which part of the
quadratic our data fall on.

On larger scales the distribution will have some downward curvature, but so
do many of the distributions claimed to follow power laws, so it is
possible that these distributions are really log-normal.  In fact, in many
cases we don't even have to restrict ourselves to a particularly small a
portion of the curve.  If $\sigma$ is large then the quadratic term in
Eq.~\eref{lnquad} will vary slowly and the curvature of the line will be
slight, so the distribution will appear to follow a power law over
relatively large portions of its range.  This situation arises commonly
when we are considering products of random numbers.

Suppose for example that we are multiplying together 100 numbers, each of
which is drawn from some distribution such that the standard deviation of
the logs is around~1---i.e.,~the numbers themselves vary up or down by
about a factor of~$\e$.  Then, by the central limit theorem, the standard
deviation for~$\ln x$ will be $\sigma\simeq10$ and $\ln x$ will have to
vary by about $\pm10$ for changes in $(\ln x)^2/\sigma^2$ to be apparent.
But such a variation in the logarithm corresponds to a variation in $x$ of
more than four orders of magnitude.  If our data span a domain smaller than
this, as many of the plots in Fig.~\ref{examples} do, then we will see a
measured distribution that looks close to power-law.  And the range will
get quickly larger as the number of numbers we are multiplying grows.

One example of a random multiplicative process might be wealth generation
by investment.  If a person invests money, for instance in the stock
market, they will get a percentage return on their investment that varies
over time.  In other words, in each period of time their investment is
multiplied by some factor which fluctuates from one period to the next.  If
the fluctuations are random and uncorrelated, then after many such periods
the value of the investment is the initial value multiplied by the product
of a large number of random numbers, and therefore should be distributed
according to a log-normal.  This could explain why the tail of the wealth
distribution, Fig.~\ref{examples}j, appears to follow a power law.

Another example is \defn{fragmentation}.  Suppose we break a stick of unit
length into two parts at a position which is a random fraction~$z$ of the
way along the stick's length.  Then we break the resulting pieces at random
again and so on.  After many breaks, the length of one of the remaining
pieces will be $\prod_i z_i$, where $z_i$ is the position of the $i$th
break.  This is a product of random numbers and thus the resulting
distribution of lengths should follow a power law over a portion of its
range.  A mechanism like this could, for instance, produce a power-law
distribution of meteors or other interplanetary rock fragments, which tend
to break up when they collide with one another, and this in turn could
produce a power-law distribution of the sizes of meteor craters similar to
the one in Fig.~\ref{examples}g.

In fact, as discussed by a number of
authors~\cite{SC97,Sornette98,Gabaix99}, random multiplication processes
can also generate perfect power-law distributions with only a slight
modification: if there is a lower bound on the value that the product of a
set of numbers is allowed to take (for example if there is a ``reflecting
boundary'' on the lower end of the range, or an additive noise term as well
as a multiplicative one) then the behaviour of the process is modified to
generate not a log-normal, but a true power law.

Finally, some processes show power-law distributions of times between
events.  The distribution of times between earthquakes and their
aftershocks is one example.  Such power-law distributions of times are
observed in critical models and in the coherent noise mechanism mentioned
above, but another possible explanation for their occurrence is a
\defn{random extremal process} or \defn{record dynamics}.  In this
mechanism we consider how often a randomly fluctuating quantity will break
its own record for the highest value recorded.  For a quantity with, say, a
Gaussian distribution, it is always in theory possible for the record to be
broken, no matter what its current value, but the more often the record is
broken the higher the record will get and the longer we will have to wait
until it is broken again.  As shown by Sibani and Littlewood~\cite{SL93},
this non-stationary process gives a distribution of waiting times between
the establishment of new records that follows a power law with
exponent~$\alpha=1$.  Interestingly, this is precisely the exponent
observed for the distribution of waiting times for aftershocks of
earthquakes.  The record dynamics has also been proposed as a model for the
lifetimes of biological taxa~\cite{SSA95}.

\section{Conclusions}
In this review I have discussed the power-law statistical distributions
seen in a wide variety of natural and man-made phenomena, from earthquakes
and solar flares to populations of cities and sales of books.  We have seen
many examples of power-law distributions in real data and seen how to
analyse those data to understand the behaviour and parameters of the
distributions.  I~have also described a number of physical mechanisms that
have been proposed to explain the occurrence of power laws.  Perhaps the
two most important of these are:
\begin{enumerate}
\item The Yule process, a rich-get-richer mechanism in which the most
populous cities or best-selling books get more inhabitants or sales in
proportion to the number they already have.  Yule and later Simon showed
mathematically that this mechanism produces what is now called the Yule
distribution, which follows a power law in its tail.
\item Critical phenomena and the associated concept of self-organized
criticality, in which a scale-factor of a system diverges, either because
we have tuned the system to a special critical point in its parameter space
or because the system automatically drives itself to that point by some
dynamical process.  The divergence can leave the system with no appropriate
scale factor to set the size of some measured quantity and as we have seen
the quantity must then follow a power law.
\end{enumerate}

The study of power-law distributions is an area in which there is
considerable current research interest.  While the mechanisms and
explanations presented here certainly offer some insight, there is much
work to be done both experimentally and theoretically before we can say we
really understand the physical processes driving these systems.  Without
doubt there are many exciting discoveries still waiting to be made.

\section*{Acknowledgements}
The author thanks Jean-Philippe Bouchaud, Petter Holme, Cris Moore, Cosma
Shalizi, Eduardo Sontag, Didier Sornette, and Erik van Nimwegen for useful
conversations and suggestions, and Lada Adamic for the web site hit data.
This work was funded in part by the National Science Foundation under grant
number DMS--0405348.

\appendix

\section{Rank/frequency plots}
\label{rfappendix}
Suppose we wish to make a plot of the cumulative distribution
function~$P(x)$ of a quantity such as, for example, the frequency with
which words appear in a body of text (Fig.~\ref{examples}a).  We start by
making a list of all the words along with their frequency of occurrence.
Now the cumulative distribution of the frequency is defined such that
$P(x)$ is the fraction of words with frequency greater than or equal
to~$x$.  Or alternatively one could simply plot the \emph{number} of words
with frequency greater than or equal to~$x$, which differs from the
fraction only in its normalization.

Now consider the most frequent word, which is ``the'' in most written
English texts.  If $x$ is the frequency with which this word occurs, then
clearly there is exactly one word with frequency greater than or equal
to~$x$, since no other word is more frequent.  Similarly, for the frequency
of the second most common word---usually ``of''---there are two words with
that frequency or greater, namely ``of'' and ``the''.  And so forth.  In
other words, if we rank the words in order, then by definition there are
$n$ words with frequency greater than or equal to that of the $n$th most
common word.  Thus the cumulative distribution $P(x)$ is simply
proportional to the rank $n$ of a word.  This means that to make a plot of
$P(x)$ all we need do is sort the words in decreasing order of frequency,
number them starting from~1, and then plot their ranks as a function of
their frequency.  Such a plot of rank against frequency was called by
Zipf~\cite{Zipf49} a \defn{rank/frequency plot}, and this name is still
sometimes used to refer to plots of the cumulative distribution of a
quantity.  Of course, many quantities we are interested in are not
frequencies---they are the sizes of earthquakes or people's personal wealth
or whatever---but nonetheless people still talk about ``rank/frequency''
plots although the name is not technically accurate.

In practice, sorting and ranking measurements and then plotting rank
against those measurements is usually the quickest way to construct a plot
of the cumulative distribution of a quantity.  All the cumulative plots in
this paper were made in this way, except for the plot of the sizes of moon
craters in Fig.~\ref{examples}g, for which the data came already in
cumulative form.

\section{Maximum likelihood estimate of exponents}
\label{mlmethod}
Consider the power-law distribution
\begin{equation}
p(x) = Cx^{-\alpha} = {\alpha-1\over x_\mathrm{min}}
                      \biggl( {x\over x_\mathrm{min}} \biggr)^{-\alpha},
\label{mlepl}
\end{equation}
where we have made use of the value of the normalization constant~$C$
calculated in Eq.~\eref{normc}.

Given a set of~$n$ values $x_i$, the probability that those values were
generated from this distribution is proportional to
\begin{equation}
P(x|\alpha) = \prod_{i=1}^n p(x_i)
            = \prod_{i=1}^n {\alpha-1\over x_\mathrm{min}}
                      \biggl( {x_i\over x_\mathrm{min}} \biggr)^{-\alpha}.
\end{equation}
This quantity is called the \defn{likelihood} of the data set.

To find the value of~$\alpha$ that best fits the data, we need to calculate
the probability~$P(\alpha|x)$ of a particular value of $\alpha$ given the
observed~$\set{x_i}$, which is related to $P(x|\alpha)$ by Bayes' law thus:
\begin{equation}
P(\alpha|x) = P(x|\alpha) {P(\alpha)\over P(x)}.
\end{equation}
The prior probability of the data $P(x)$ is fixed since $x$ itself is
fixed---$x$~is equal to the particular set of observations we actually made
and does not vary in the calculation---and it is usually assumed, in the
absence of any information to the contrary, that the prior probability of
the exponent $P(\alpha)$ is uniform, i.e.,~a constant independent
of~$\alpha$.  Thus $P(\alpha|x) \propto P(x|\alpha)$.  For convenience we
typically work with the logarithm of~$P(\alpha|x)$, which, to within an
additive constant, is equal to the log of the likelihood, denoted~$\cL$ and
given by
\begin{eqnarray}
\cL &=& \ln P(x|\alpha) 
     = \sum_{i=1}^n \biggl[ \ln(\alpha-1) - \ln x_\mathrm{min}
       - \alpha \ln {x_i\over x_\mathrm{min}} \biggr] \nonumber\\
    &=& n\ln(\alpha-1) - n\ln x_\mathrm{min}
        - \alpha \sum_{i=1}^n \ln {x_i\over x_\mathrm{min}}.
\label{loglikelihood}
\end{eqnarray}
Now we calculate the most likely value of $\alpha$ by maximizing the
likelihood with respect to~$\alpha$, which is the same as maximizing the
log likelihood, since the logarithm is a monotonic increasing function.
Setting $\partial\cL/\partial\alpha=0$, we find
\begin{equation}
{n\over\alpha-1} -  \sum_{i=1}^n \ln {x_i\over x_\mathrm{min}} = 0,
\label{maxlike}
\end{equation}
or
\begin{equation}
\alpha = 1 + n \biggl[ \sum_i \ln {x_i\over x_\mathrm{min}} \biggr]^{-1}.
\label{appmle}
\end{equation}

We also wish to know what the expected error is on our value of~$\alpha$.
We can estimate this from the width of the maximum of the likelihood as a
function of~$\alpha$.  Taking the exponential of Eq.~\eref{loglikelihood},
we find that that the likelihood has the form
\begin{equation}
P(x|\alpha) = a \e^{-b\alpha} (\alpha-1)^n,
\end{equation}
where $b=\sum_{i=1}^n \ln (x_i/x_\mathrm{min})$ and $a$ is an unimportant
normalizing constant.  Assuming that $\alpha>1$ so that the
distribution~\eref{mlepl} is normalizable, the mean and mean square
of~$\alpha$ in this distribution are given by
\begin{eqnarray}
\av{\alpha} &=& {\int_1^\infty \e^{-b\alpha}
                (\alpha-1)^n \alpha \>\d\alpha\over
                \int_1^\infty \e^{-b\alpha} (\alpha-1)^n \>\d\alpha}\nonumber\\
            &=& {\e^{-b} b^{-2-n} (n+1+b) \Gamma(n+1)\over
                \e^{-b} b^{-1-n} \Gamma(n+1)} \nonumber\\
            &=& {n+1+b\over b}
\end{eqnarray}
and
\begin{eqnarray}
\av{\alpha^2} &=& {\int_1^\infty \e^{-b\alpha}
                (\alpha-1)^n \alpha^2 \>\d\alpha\over
                \int_1^\infty \e^{-b\alpha} (\alpha-1)^n \>\d\alpha}
                \nonumber\\
            &=& {\e^{-b} b^{-3-n} (n^2+3n+b^2+2b+2nb+2) \Gamma(n+1)\over
                \e^{-b} b^{-1-n} \Gamma(n+1)} \nonumber\\
            &=& {n^2+3n+b^2+2b+2nb+2\over b^2},
\end{eqnarray}
where $\Gamma(x)$ is the $\Gamma$-function of Eq.~\eref{defsgamma}.  Then
the variance of~$\alpha$ is
\begin{eqnarray}
\sigma^2 &=& \av{\alpha^2} - \av{\alpha}^2 \nonumber\\
  &=& {n^2+3n+b^2+2b+2nb+2\over b^2} - {(n+1+b)^2\over b^2} \nonumber\\
  &=& {n+1\over b^2},
\end{eqnarray}
and the error on~$\alpha$ is
\begin{equation}
\sigma = {\sqrt{n+1}\over b}
  = \sqrt{n+1} \biggl[ \sum_i \ln {x_i\over x_\mathrm{min}} \biggr]^{-1}.
\end{equation}
In most cases we will have $n\gg1$ and it is safe to approximate $n+1$
by~$n$, giving
\begin{equation}
\sigma
  = \sqrt{n} \biggl[ \sum_i \ln {x_i\over x_\mathrm{min}} \biggr]^{-1}
  = {\alpha-1\over\sqrt{n}},
\end{equation}
where $\alpha$ in this expression is the maximum likelihood estimate from
Eq.~\eref{appmle}.

\end{document}